\documentclass[aps, prl, reprint, superscriptaddress]{revtex4-2} 
\usepackage[utf8]{inputenc}
\usepackage[sort&compress]{natbib}
\usepackage{ulem}
\usepackage{overpic}
\usepackage{bm}
\usepackage{times}
\usepackage{amssymb,amsbsy,amsmath,amsfonts}
\usepackage{graphicx}
\usepackage{color}
\usepackage{booktabs}
\usepackage{makecell}

\usepackage{rotating}
\usepackage{srcltx}
\usepackage{slashed}
\usepackage{subfigure}
\usepackage{multirow}
\usepackage{verbatim}
\usepackage{hyperref}
\usepackage{tabularx}

\DeclareUnicodeCharacter{3002}{HEREHEREHERE}
\newcommand{\dd}{\mathrm{d}}
\newcommand{\ii}{\mathrm{i}}
\newcommand{\ee}{\mathrm{e}} 

\newcolumntype{C}{>{\centering\arraybackslash}X}

\begin{document}

\title{Deformation effects on reaction observables of beryllium nuclei from \textit{ab initio} densities}

\author{Qi Lu}
\affiliation{School of Physics, Beihang University, Beijing 102206, China}

\author{Rui-Feng Tian}
\affiliation{School of Physics, Beihang University, Beijing 102206, China}
\affiliation{Peng Huanwu Collaborative Center for Research and Education, International Institute for Interdisciplinary and Frontiers, Beihang University, Beijing 100191, China}

\author{Shi-Sheng Zhang}\email[Corresponding author:]{zss76@buaa.edu.cn}
\affiliation{School of Physics, Beihang University, Beijing 102206, China}

\author{Ulf-G. Meißner}
\email{meissner@hiskp.uni-bonn.de}
\affiliation{Helmholtz-Institut für Strahlen- und Kernphysik, Bethe Center for Theoretical Physics
and Cluster of Excelence -- ``Color meets Flavor'', , Universität Bonn, D-53115 Bonn, Germany}
\affiliation{Institute for Advanced Simulation (IAS-4), Forschungszentrum Jülich, D-52425 Jülich, Germany}
\affiliation{Peng Huanwu Collaborative Center for Research and Education, International Institute for Interdisciplinary and Frontiers, Beihang University, Beijing 100191, China}

\author{Shihang Shen}\email[Corresponding author:]{sshen@buaa.edu.cn}
\affiliation{Peng Huanwu Collaborative Center for Research and Education, International Institute for Interdisciplinary and Frontiers, Beihang University, Beijing 100191, China}
\affiliation{School of Physics, Beihang University, Beijing 102206, China}

\date{\today}

\begin{abstract}
We combine three-dimensional intrinsic densities from \textit{ab initio} nuclear lattice effective field theory with a deformed Glauber model to study high-energy reactions of $^{7\text{--}12}$Be.
To connect the correlated many-body configurations to the core-plus-neutron reaction formalism without imposing a single-particle orbital, we introduce a configuration-resolved prescription that identifies the spatially outermost valence neutron after the two-cluster decomposition.
For Be projectiles on $^{12}$C and $^{9}$Be targets at 790~MeV/$A$, explicit orientation averaging lowers the calculated reaction cross section of $^{11}$Be by up to approximately 50~mb relative to a calculation with the spherically averaged density.
The deformed calculation reproduces the pronounced increase from $^{10}$Be to the established one-neutron halo nucleus $^{11}$Be for both targets.
We further calculate the momentum distribution of the fragments after the one-neutron removal reaction of $^{11}$Be$+^{9}$Be 
, finding good agreement in shape with the measurement at 63~MeV/$A$ and providing a prediction at 790~MeV/$A$.
These results quantify how intrinsic deformation and weak binding are transmitted from microscopic many-body densities to
reaction observables.
\end{abstract}

\maketitle

{\it Introduction---}
Radioactive ion beams enable the study of exotic nuclei far from the valley of stability. Such nuclei commonly exhibit novel phenomena, such as nuclear halos~\cite{Tanihata2013PPNP} and nucleon emission. Because of nuclear deformation and the population of bound and resonant states around the Fermi surface in loosely bound systems, new magic numbers and islands of inversion~\cite{Nakamura2009PRL_262501,Kobayashi2014PRL_242501} can occur. These features played an important role in identifying the medium-mass halo nuclei $^{31}$Ne and $^{37}$Mg. A long-standing challenge is the so-called quenching of spectroscopic strength, which characterizes systematic deviations between theoretical predictions and measured nucleon-removal cross sections~\cite{Tostevin2014PRC,Tostevin2021PRC}.

Experimentally, reaction observables are commonly used to identify dilute halo distributions through intermediate- and high-energy reactions. Examples include a sudden increase of reaction or interaction cross sections along an isotopic chain, soft $E1$ Coulomb excitation, and narrow momentum distributions following valence neutron removal~\cite{Tanihata1985PRL,Fukuda1991PLB,Tanihata1988PLB,Nakamura1994PLB,Fauerbach1997PRC,hansen1987neutron,jensen2004structure,Tanihata2013PPNP}.
Because the experimental identification of halos becomes increasingly difficult in heavier mass regions, searches for new light halo candidates have attracted renewed interest.
The established one-neutron halo nucleus $^{11}$Be~\cite{Fukuda1991PLB} and the more recently identified halo features of $^{17}$B~\cite{Yang2021PRL} provide stringent benchmarks.
In particular, $^{11}$Be is a paradigmatic one-neutron ($1n$) $s$-wave halo nucleus~\cite{Fukuda1991PLB}. Its microscopic description is nevertheless complicated by the pronounced clustering, deformation, and configuration mixing characteristic of light beryllium isotopes~\cite{Freer2018RMP,Shen:2024qzi}.

The theoretical interpretation of such experiments relies largely on Glauber theory, which has become the standard framework for high-energy nucleon-nucleus and nucleus-nucleus collisions. For several decades, the Glauber model has been extensively used to extract nuclear radii and density distributions from reaction measurements. Connections from microscopic structure to reaction observables have also been established for deformed medium-mass halo nuclei. For example, the $1n$ $p$-wave halo nucleus $^{37}$Mg~\cite{Kobayashi2014PRL_242501,Takechi2014PRC_061305} has been studied by combining deformed relativistic Hartree-Bogoliubov theory in continuum~\cite{Zhou2010PRC(R),Li2012PRC,Zhang2020PRC,Pan2022PRC} with Glauber reaction calculations~\cite{An2024}.

More recently, Glauber theory combined with realistic \textit{ab initio} many-body wave functions has achieved quantitative descriptions of reaction observables~\cite{Horiuchi2026PRL,Horiuchi2026PRC}. These developments demonstrate that high-energy measurements can be connected directly to microscopic nuclear structure calculations and establish Glauber theory as a bridge between nuclear structure and reactions.

While reaction observables have traditionally been used to determine nuclear sizes and density distributions, considerably less attention has been paid to their sensitivity to intrinsic nuclear shape. Nuclear deformation is a fundamental manifestation of many-body dynamics and has conventionally been studied through low-energy spectroscopy and electromagnetic transitions. Recently, collective-flow analyses in ultrarelativistic heavy-ion collisions have demonstrated that high-energy collisions can provide information on intrinsic nuclear 
deformation~\cite{STAR:2024wgy,Giacalone:2024luz,ATLAS:2025nnt,ALICE:2025luc}. This raises a natural question: can high-energy reaction observables also serve as quantitative probes of nuclear shape?

Answering this question requires both a microscopic description of deformation and a reaction framework that preserves orientation-dependent structure information. Recent \textit{ab initio} nuclear lattice effective field theory (NLEFT) calculations of the beryllium isotopes have revealed clustering, halo formation, and pronounced deformation while providing microscopic three-dimensional matter densities~\cite{Shen:2024qzi,Shen:2026liw}. In many applications to halo nuclei, the Glauber model has used spherically averaged or phenomenological densities, whereas treatments of deformation have mainly employed few-body or mean-field structure inputs~\cite{Christley1999PRC,abu2003cross,Horiuchi2012PRC,Urata2012,Simpson2012,Hong2017}. Deformed extensions of Glauber theory can incorporate orientation-dependent densities, but their applications have remained limited in part because realistic microscopic intrinsic densities were unavailable.

In this Letter, we combine, for the first time, \textit{ab initio} intrinsic densities from NLEFT with a deformed Glauber framework to investigate the influence of nuclear deformation on 
reaction observables. Using microscopic structure inputs for the beryllium isotopes, we study elastic scattering, reaction cross sections, and $1n$-removal observables. We show that intrinsic deformation significantly modifies the calculated reaction cross sections relative to conventional spherical calculations and improves agreement with the available measurements.

{\it Ab initio structure inputs from NLEFT}---
The structure inputs for the Glauber calculations are taken from our recent
\textit{ab initio}  
NLEFT study of
$^{7-12}$Be~\cite{Shen:2024qzi,Shen:2026liw}.
We employ the wavefunction matching method~\cite{Elhatisari:2022zrb}
to mitigate the Monte Carlo sign problem associated with realistic
N$^3$LO chiral interactions. A unitary transformation is applied to the
original Hamiltonian to generate a Hamiltonian whose
low-energy two-body wave functions closely match those of a sign-friendly
Hamiltonian at short distances while remaining unchanged at larger
separations, leading to a rapidly convergent expansion in the difference
between the two Hamiltonians.

Ground-state observables are computed through Euclidean-time projection,
\begin{equation}
\langle O\rangle =
\lim_{\tau\rightarrow\infty}
\frac{
\langle\Psi_0|M^{L_t/2} O M^{L_t/2}|\Psi_0\rangle
}{
\langle\Psi_0|M^{L_t}|\Psi_0\rangle
},
\end{equation}
where $O$ denotes the operator of interest,
$|\Psi_0\rangle$ is the initial trial state,
$M=:e^{-a_t H}:$ is the normal-ordered transfer matrix associated with
the Hamiltonian $H$,
$a_t$ is the temporal lattice spacing,
$L_t$ is the number of Euclidean-time steps, and
$\tau=L_t a_t$ is the total projection time.
The resulting NLEFT calculations reproduce the spectra, electromagnetic
observables, and deformation properties of the beryllium isotopes
without phenomenological adjustments~\cite{Shen:2024qzi,Shen:2026liw}.

To obtain density distributions, we employ the pinhole algorithm
and its perturbative extension~\cite{Elhatisari:2017eno,Lu:2018bat},
which sample the complete $A$-body nucleon configurations from the
projected many-body wave function.
After removing the center-of-mass motion, each configuration is rotated into
an intrinsic frame defined by the two most compact $2p$-$2n$ clusters,
corresponding to the $\alpha$-$\alpha$ symmetry axis of the beryllium isotopes.
The intrinsic densities are then constructed by
accumulating several million pinhole configurations.
These fully microscopic three-dimensional intrinsic densities are used
directly as inputs to the deformed Glauber calculations below.

{\it The deformed Glauber model}---
The reaction dynamics is described in the eikonal Glauber
framework~\cite{glauber1959lectures,abu2003cross}, using microscopic
structure inputs from the NLEFT calculations. For a projectile specified by
its intrinsic matter density $\rho_P(\bm r;\Omega)$, where $\Omega$ denotes
the orientation of the intrinsic symmetry axis with respect to the beam axis,
the projectile-target $S$ matrix is written as
\begin{equation}
S_{PT}(\bm b;\Omega)=\exp \!\left[\ii \chi_{PT}(\bm b;\Omega)\right],
\end{equation}
where $\bm b$ is the transverse impact parameter. In the optical-limit
approximation (OLA), the phase-shift function $\chi_{PT}$ is obtained by folding projectile
 density with the target density $\rho_T(\bm r')$ through the profile function $\Gamma_{NN}$,
\begin{align}
\ii\chi_{PT}(\bm b;\Omega)&=
-\int \dd\bm r \int \dd\bm r'\,
\nonumber\\
& \times\rho_P(\bm r;\Omega)\,\rho_T(\bm r')\,
\Gamma_{NN}\!\left(\bm b+\bm s-\bm s'\right),
\label{eq:chi_profile}
\end{align}
where $\bm s$ and $\bm s'$ are the transverse
components of $\bm r$ and $\bm r'$, respectively. 
In the commonly used spherical Glauber model, the density is angle-averaged
so that $S_{PT}$ becomes orientation-independent. 
In the deformed case, $S_{PT}(\bm b;\Omega)$ becomes orientation-dependent.
Thus, for unpolarized projectile beams, the reaction cross section is averaged over all orientations~\cite{Hong2017} as below,
\begin{equation}
\sigma_R(P+T)=\frac{1}{4\pi}\int \dd\Omega \int \dd \bm b\,
\left[1-\left|S_{PT}(\bm b;\Omega)\right|^2\right].
\label{eq:sigmaR}
\end{equation}

Because the correlated NLEFT state does not assign nucleons to single-particle orbitals, we use a configuration-based prescription for odd-$A$ Be projectiles.
Among the valence neutrons identified by the grouping procedure above, we select the one farthest from the total center of mass and the remaining $A-1$ nucleons define the residual core.
Accumulating its coordinate relative to the core center-of-mass yields an orientation-dependent effective valence neutron distribution.
This outermost-neutron coordinate is retained explicitly in the core-plus-neutron Glauber calculation~\cite{abu2003cross}, together with the orientation dependence of the core-target $S$ matrix, so that the $S$ matrix in Eq.~\eqref{eq:sigmaR} is then
replaced by
\begin{equation}
S_{c+n,T}(\bm b_c;\Omega)=
\left\langle \phi_\Omega \left|
S_{cT}(\bm b_c;\Omega)S_{nT}(\bm b_c+\bm s)
\right|\phi_\Omega \right\rangle ,
\label{eq:halo_S}
\end{equation}
where $\phi_\Omega$ is the valence neutron relative wave function, 
and $S_{cT}$ and $S_{nT}$ are
the core-target and neutron-target $S$ matrices, respectively. 
Therefore, unlike the OLA, which washes out core-valence correlations, we integrate the extended valence neutron coordinate exactly.
Similar to Ref.~\cite{abu2003cross}, the orientation-dependent elastic scattering amplitude $F(\bm q;\Omega)$ in our calculations has the form 
\begin{align}
F(\bm q;\Omega)&=\ee^{\ii\chi_s}
\bigg[
F_{\rm Coul}(\bm q)
\nonumber\\
&\quad+
\frac{\ii K}{2\pi}\int \dd\bm b\,
\ee^{-\ii\bm q\cdot\bm b+2\ii\eta\ln(Kb)}
\nonumber\\
&\quad\times
\left(1-
\left\langle \phi_\Omega \left|
S_{cT}(\bm b_c;\Omega)S_{nT}(\bm b_n)
\right|\phi_\Omega\right\rangle
\right)
\bigg],
\label{eq:elastic_amp}
\end{align}
where \(F_{\rm Coul}\) and \(\chi_s\) are the Coulomb elastic-scattering amplitude and screening phase-shift, respectively. 
Here, $\bm q$ denotes the momentum transfer,
$\eta$ is the Sommerfeld parameter,  
and $\bm b_n=\bm b_c+\bm s$ refers to the impact factor of the valence neutron.

We also calculate the longitudinal
momentum distribution of the residues from the $1n$ removal reaction as another observable.
With $\bm r=(\bm s,z)$, the momentum distribution
reads~\cite{abu2003cross,Hong2017}
\begin{align}
\frac{\dd\sigma^{\rm inel}_{\rm -N}}{\dd p_z}
&=\frac{1}{2\pi\hbar}\,\frac{1}{4\pi}\int 
\dd\Omega \int \dd \bm{b_n}\,
\left[1-|S_n(\bm b_n)|^2\right]
\nonumber\\
&\times \int \dd \bm s \,
\left|\int \dd z\,\ee^{-\ii p_z z/\hbar}\,
S_c(\bm b_c;\Omega)\,\phi(\bm r;\Omega)\right|^2 .
\label{eq:dsig_inel}
\end{align}

\begin{figure}[htbp]
    \centering
    \includegraphics[width=8cm]{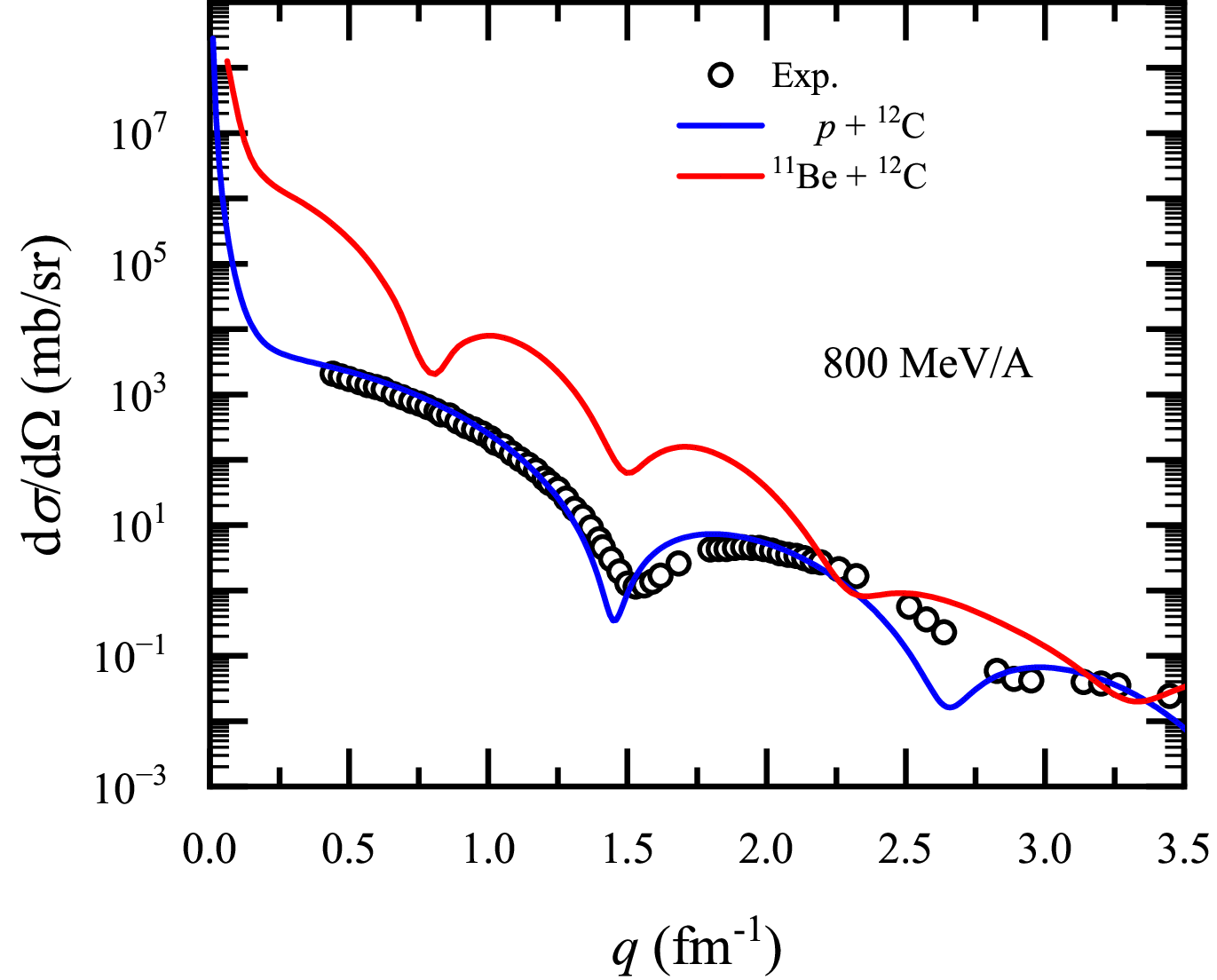}
    \caption{
    Elastic differential cross sections for $p+{}^{12}$C at 800~MeV (blue curve) and the prediction for $^{11}$Be$+{}^{12}$C at 800~MeV/$A$ with the deformed NLEFT density (red curve), plotted as functions of momentum transfer. The open circles are experimental data for $p+{}^{12}$C and are taken from Ref.~\cite{Blanpied1981PRC}.}
    \label{fig:diffElaCS}
\end{figure}

{\it Results and discussion}---
Figure~\ref{fig:diffElaCS} shows the elastic differential cross sections for $p+{}^{12}$C at 800~MeV and $^{11}$Be$+{}^{12}$C at 800~MeV/$A$.
The calculated $p+{}^{12}$C cross section follows the measured diffraction pattern through approximately the second minimum.
This agreement tests the $^{12}$C density, the nucleon-nucleon profile function, and the elastic-scattering implementation.
The recent full Glauber calculations with a VMC $^{12}$C wave function provide a stringent assessment of the OLA and likewise describe the proton data well in this momentum-transfer region~\cite{Horiuchi2026PRL,Horiuchi2026PRC}.
The red curve is our prediction for $^{11}$Be$+{}^{12}$C.
Compared with the $p+^{12}$C reaction, the calculated
$^{11}$Be$+^{12}$C cross section exhibits a distinct diffraction
pattern, with the first minimum shifted toward lower $q$.
Because the nuclear elastic amplitude is the two-dimensional Fourier
transform of the profile function, this shift reflects the
larger transverse interaction radius of the $^{11}$Be$+^{12}$C
system which is consistent with the spatially extended valence neutron
distribution in $^{11}$Be.

\begin{figure}[htbp]
    \centering
    \includegraphics[width=8cm]{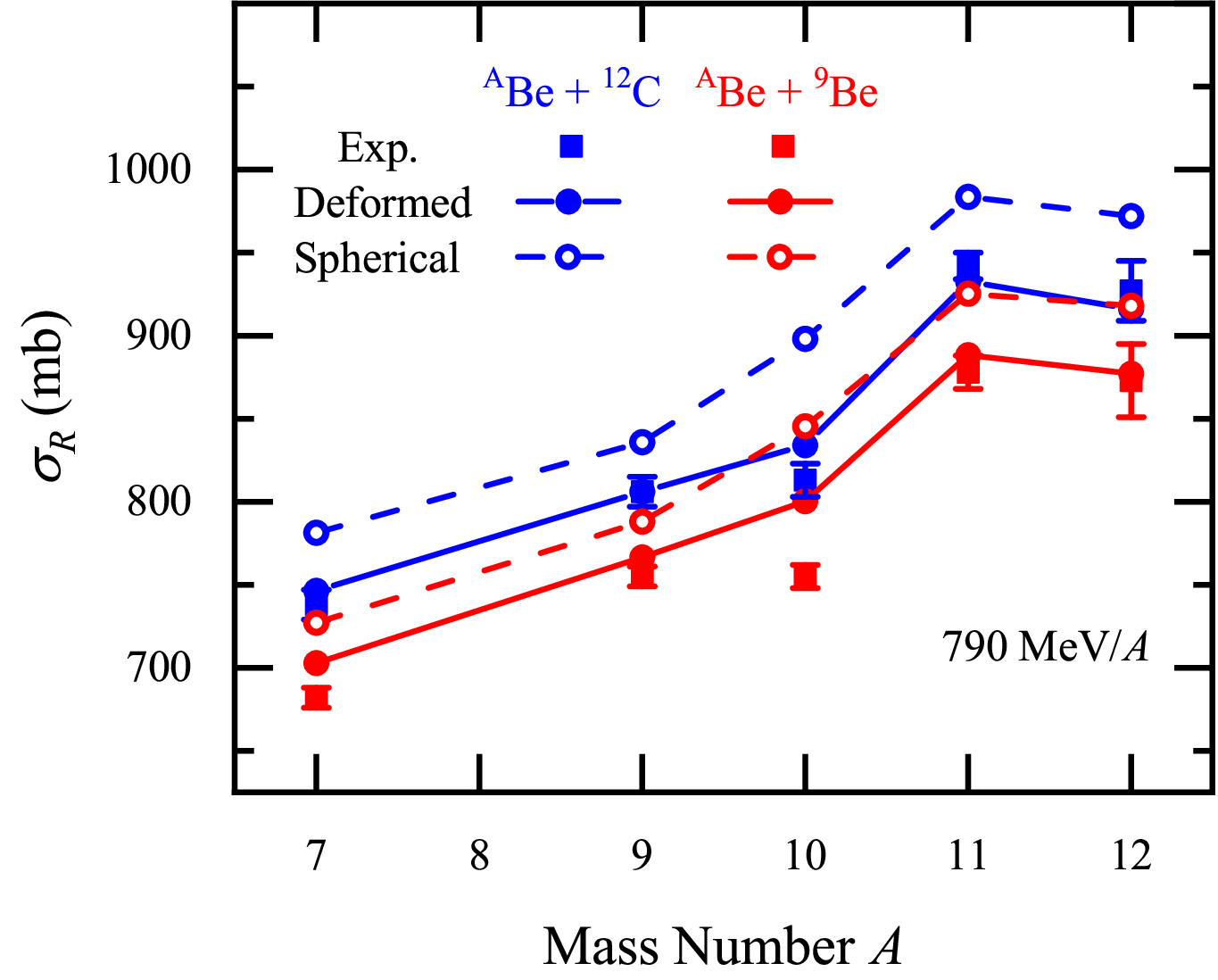}
    \caption{Calculated reaction cross sections $\sigma_R$ for $^{7\text{--}12}$Be projectiles on $^{12}$C (blue) and $^{9}$Be (red) targets at 790~MeV/$A$. Solid circles show orientation-averaged calculations with the intrinsic NLEFT densities, open circles show calculations with the corresponding spherically averaged densities. The solid squares are measured interaction cross sections from Refs.~\cite{Tanihata1985PRL,Tanihata1988PLB}.}
    \label{fig:reacCS}
\end{figure}

We next examine the isotope dependence of the reaction cross sections. Figure ~\ref{fig:reacCS} compares calculations for $^{7\text{--}12}$Be projectiles on $^{12}$C and $^{9}$Be targets at 790~MeV/$A$.
The measured quantities shown by squares are the interaction cross sections, whereas the calculations give total reaction cross sections.
The difference of the two cross sections is small at these energies~\cite{Kohama2008PRC,Horiuchi2026PRL,Horiuchi2026PRC}.

The open circles use the spherically averaged NLEFT densities, while the solid circles retain each intrinsic density and average the reaction probability over orientations.
Retaining the deformation systematically reduces the calculated reaction cross sections, with the reduction reaching approximately 50~mb for $^{11}$Be.
The reduced cross sections provide a markedly better description across the Be isotopic chain. For both targets, nearly all deformed results lie close to the measured values, with $^{10}$Be being the main exception.
This comparison is controlled because the projectile configurations, target densities, nucleon-nucleon profile function, and all other reaction inputs are unchanged; the only difference is whether the projectile density is averaged over angles before the reaction calculation or its orientation dependence is retained explicitly.
The difference between the two results therefore directly quantifies the deformation effect within the present Glauber model.

The most prominent isotope-dependent consequence appears between $^{10}$Be and $^{11}$Be.
In the spherical calculation, this increase is moderate, whereas retaining the intrinsic deformation produces a pronounced step for both targets, in agreement with the experimental trend associated with the spatially extended halo structure of $^{11}$Be.
Its occurrence on both carbon and beryllium targets indicates that the enhancement is driven primarily by the projectile structure rather than by a target-specific effect.

This reduction is consistent with previous deformed Glauber analyses~\cite{Christley1999PRC,Hong2017,Rashdan2019}.
The present framework retains the orientation-dependent core density and, for odd-$A$ projectiles, treats the valence neutron coordinate explicitly through Eq.~\eqref{eq:halo_S}
and therefore already includes the dominant
contribution beyond the OLA for a weakly bound projectile~\cite{abu2003cross} .
It does not, however, include two-body-density contributions within the core and target.
Horiuchi \textit{et al.} showed that such second-cumulant contributions are important for nucleus-nucleus scattering, particularly at larger momentum transfers, and bring the result close to the full Glauber calculation~\cite{Horiuchi2026PRL,Horiuchi2026PRC}.
The present approach thus provides a geometry-resolved extension of 
the spherical OLA, 
complementary to correlation-resolved full Glauber calculations.

\begin{figure}[htbp]
    \centering
    \includegraphics[width=8cm]{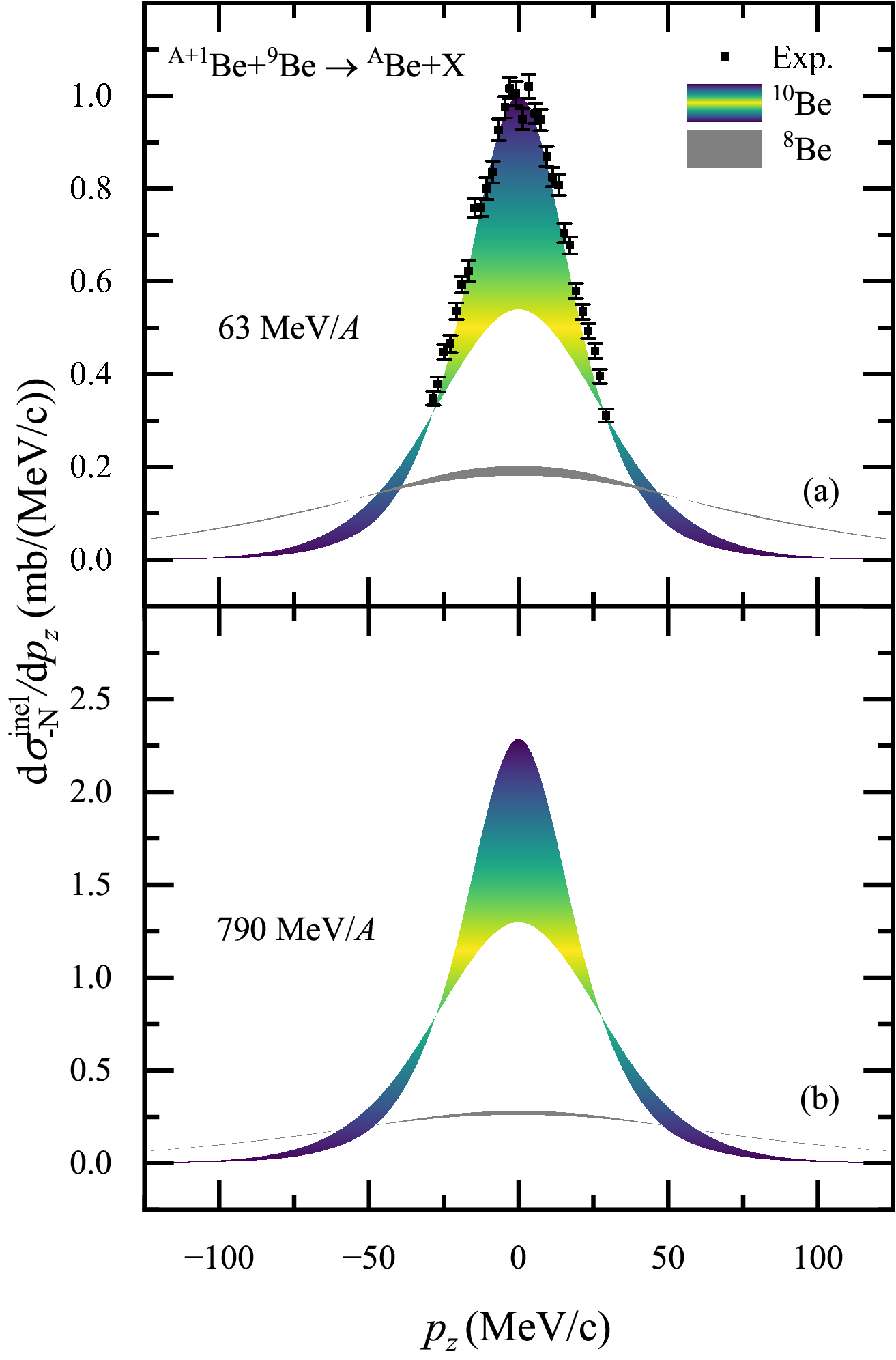}
    \caption{
    Longitudinal momentum distributions $\mathrm{d}\sigma^{\rm inel}_{\rm -N}/\mathrm{d}p_z$ of the $^{A}$Be residues after $1n$ removal reactions of $^{A+1}$Be on $^{9}$Be target  at (a) 63~MeV/$A$ and (b) 790~MeV/$A$.
    The bands represent the sensitivity to $1n$ separation energy $S_n$ used in the wave function of valence neutron, and the black squares in panel (a) are the experimental data from Ref.~\cite{Kelley1995PRL}.
    }
    \label{fig:mom}
\end{figure}

In Fig.~\ref{fig:mom}, we exhibit the longitudinal momentum distributions
$\mathrm{d}\sigma^{\rm inel}_{\rm -N}/\mathrm{d}p_z$ of the residues
from $1n$ removal reactions
$^{A+1}$Be$+^{9}$Be $\to$ $^{A}$Be$+X$ at 63~MeV/$A$ and
790~MeV/$A$, shown in panels~(a) and~(b) respectively. 
The bands quantify the sensitivity to the one-neutron separation energy used to continue the valence neutron wave function beyond the finite NLEFT volume.
The NLEFT calculations of Ref.~\cite{Shen:2024qzi} used a periodic box of length 13.2~fm. For $r>r_m=6$~fm, we match the sampled valence neutron density to the asymptotic form~\cite{Liu2003PRC}
\begin{equation}
\rho_v(r) \simeq b_{l j}^2 \frac{W_{-\eta, l+1/2}^2(2 \kappa r)}{r^2}, \qquad \kappa = \frac{\sqrt{2\mu S_n}}{\hbar},
\end{equation}
where $\mu$ is the core-neutron reduced mass and $b_{l j}$ is fixed by matching at $r_m$.
Because the momentum distribution is sensitive to the asymptotic wave function~\cite{Esbensen1996PRC,Hebborn2019PRC}, we use both the experimental and NLEFT values of $S_n$ as a statistical uncertainty.
For $^{11}$Be, the colored band spans $S_n^{\rm exp}=0.5$~MeV to $S_n^{\rm NLEFT}=1.9$~MeV; for $^{9}$Be, the gray band spans $S_n^{\rm NLEFT}=0.9$~MeV to $S_n^{\rm exp}=1.7$~MeV. 

Across these choices, the $^{10}$Be residue distribution remains substantially narrower than the $^{8}$Be distribution. 
In panel~(a), the calculated $^{10}$Be residues distribution at 63~MeV/$A$ agrees well in shape with the data (black squares), while panel~(b) provides the corresponding prediction at 790~MeV/$A$.
This follows from the more extended core-neutron wave function in $^{11}$Be.
The momentum distribution therefore provides a complementary probe of the asymptotic halo tail, whereas the isotope dependence of the integrated reaction cross section is also sensitive to the overall radius and intrinsic deformation.

{\it Summary}---
We have combined \textit{ab initio} NLEFT intrinsic densities with a deformed Glauber model to study intermediate- and high-energy reactions of $^{7\text{--}12}$Be.
The calculated $p$+$^{12}$C elastic cross section reproduces the measured diffraction pattern, and we predict the pattern for $^{11}$Be+$^{12}$C.
Retaining orientation-dependent intrinsic densities reduces reaction cross sections by up to $\sim\!50$~mb for $^{11}$Be relative to spherical averaging, yielding a pronounced increase from $^{10}$Be to $^{11}$Be that agrees with the halo signature for both targets.
The longitudinal momentum distribution of $^{10}$Be residues after one-neutron removal from $^{11}$Be matches the shape measured at 63~MeV/$A$ and is substantially narrower than the $^{8}$Be distribution; its sensitivity to the separation energy reflects the asymptotic halo tail.
These results demonstrate that intermediate- and high-energy reactions provide a complementary window on intrinsic nuclear structure, allowing deformation and weak binding predicted by microscopic many-body theories to be directly tested through reaction observables.

{\it Acknowledgements}---
This work was partly supported by the National Natural Science Foundation of China (Grants Nos.~12575122, 12175010, 12305125, U2541242, and 12435007), the National Key Laboratory of Neutron Science and Technology (Grant No.~NST202401016), and the Sichuan Science and Technology Program (Grant No.~2024NSFSC1356).
We gratefully acknowledge the Computational resources provided by the HPC platform of Beihang University.
The work of UGM was supported in part by the European
Research Council (ERC) under the European Union's Horizon 2020 research
and innovation programme (EXOTIC, grant agreement No. 101018170),
and by the CAS President's International Fellowship Initiative (PIFI) (Grant No.~2025PD0022).

\bibliography{reference}

\begin{thebibliography}{45}%
\makeatletter
\providecommand \@ifxundefined [1]{%
 \@ifx{#1\undefined}
}%
\providecommand \@ifnum [1]{%
 \ifnum #1\expandafter \@firstoftwo
 \else \expandafter \@secondoftwo
 \fi
}%
\providecommand \@ifx [1]{%
 \ifx #1\expandafter \@firstoftwo
 \else \expandafter \@secondoftwo
 \fi
}%
\providecommand \natexlab [1]{#1}%
\providecommand \enquote  [1]{``#1''}%
\providecommand \bibnamefont  [1]{#1}%
\providecommand \bibfnamefont [1]{#1}%
\providecommand \citenamefont [1]{#1}%
\providecommand \href@noop [0]{\@secondoftwo}%
\providecommand \href [0]{\begingroup \@sanitize@url \@href}%
\providecommand \@href[1]{\@@startlink{#1}\@@href}%
\providecommand \@@href[1]{\endgroup#1\@@endlink}%
\providecommand \@sanitize@url [0]{\catcode `\\12\catcode `\$12\catcode
  `\&12\catcode `\#12\catcode `\^12\catcode `\_12\catcode `\%12\relax}%
\providecommand \@@startlink[1]{}%
\providecommand \@@endlink[0]{}%
\providecommand \url  [0]{\begingroup\@sanitize@url \@url }%
\providecommand \@url [1]{\endgroup\@href {#1}{\urlprefix }}%
\providecommand \urlprefix  [0]{URL }%
\providecommand \Eprint [0]{\href }%
\providecommand \doibase [0]{https://doi.org/}%
\providecommand \selectlanguage [0]{\@gobble}%
\providecommand \bibinfo  [0]{\@secondoftwo}%
\providecommand \bibfield  [0]{\@secondoftwo}%
\providecommand \translation [1]{[#1]}%
\providecommand \BibitemOpen [0]{}%
\providecommand \bibitemStop [0]{}%
\providecommand \bibitemNoStop [0]{.\EOS\space}%
\providecommand \EOS [0]{\spacefactor3000\relax}%
\providecommand \BibitemShut  [1]{\csname bibitem#1\endcsname}%
\let\auto@bib@innerbib\@empty
\bibitem [{\citenamefont {Tanihata}\ \emph {et~al.}(2013)\citenamefont
  {Tanihata}, \citenamefont {Savajols},\ and\ \citenamefont
  {Kanungo}}]{Tanihata2013PPNP}%
  \BibitemOpen
  \bibfield  {author} {\bibinfo {author} {\bibfnamefont {I.}~\bibnamefont
  {Tanihata}}, \bibinfo {author} {\bibfnamefont {H.}~\bibnamefont {Savajols}},\
  and\ \bibinfo {author} {\bibfnamefont {R.}~\bibnamefont {Kanungo}},\ }\href
  {https://doi.org/10.1016/j.ppnp.2012.07.001} {\bibfield  {journal} {\bibinfo
  {journal} {Prog. Part. Nucl. Phys.}\ }\textbf {\bibinfo {volume} {68}},\
  \bibinfo {pages} {215} (\bibinfo {year} {2013})}\BibitemShut {NoStop}%
\bibitem [{\citenamefont {Nakamura}\ \emph {et~al.}(2009)\citenamefont
  {Nakamura}, \citenamefont {Kobayashi}, \citenamefont {Kondo}, \citenamefont
  {Satou}, \citenamefont {Aoi}, \citenamefont {Baba}, \citenamefont {Deguchi},
  \citenamefont {Fukuda}, \citenamefont {Gibelin}, \citenamefont {Inabe},
  \citenamefont {Ishihara}, \citenamefont {Kameda}, \citenamefont {Kawada},
  \citenamefont {Kubo}, \citenamefont {Kusaka}, \citenamefont {Mengoni},
  \citenamefont {Motobayashi}, \citenamefont {Ohnishi}, \citenamefont {Ohtake},
  \citenamefont {Orr}, \citenamefont {Otsu}, \citenamefont {Otsuka},
  \citenamefont {Saito}, \citenamefont {Sakurai}, \citenamefont {Shimoura},
  \citenamefont {Sumikama}, \citenamefont {Takeda}, \citenamefont {Takeshita},
  \citenamefont {Takechi}, \citenamefont {Takeuchi}, \citenamefont {Tanaka},
  \citenamefont {Tanaka}, \citenamefont {Tanaka}, \citenamefont {Togano},
  \citenamefont {Utsuno}, \citenamefont {Yoneda}, \citenamefont {Yoshida},\
  and\ \citenamefont {Yoshida}}]{Nakamura2009PRL_262501}%
  \BibitemOpen
  \bibfield  {author} {\bibinfo {author} {\bibfnamefont {T.}~\bibnamefont
  {Nakamura}}, \bibinfo {author} {\bibfnamefont {N.}~\bibnamefont {Kobayashi}},
  \bibinfo {author} {\bibfnamefont {Y.}~\bibnamefont {Kondo}}, \bibinfo
  {author} {\bibfnamefont {Y.}~\bibnamefont {Satou}}, \bibinfo {author}
  {\bibfnamefont {N.}~\bibnamefont {Aoi}}, \bibinfo {author} {\bibfnamefont
  {H.}~\bibnamefont {Baba}}, \bibinfo {author} {\bibfnamefont {S.}~\bibnamefont
  {Deguchi}}, \bibinfo {author} {\bibfnamefont {N.}~\bibnamefont {Fukuda}},
  \bibinfo {author} {\bibfnamefont {J.}~\bibnamefont {Gibelin}}, \bibinfo
  {author} {\bibfnamefont {N.}~\bibnamefont {Inabe}}, \bibinfo {author}
  {\bibfnamefont {M.}~\bibnamefont {Ishihara}}, \bibinfo {author}
  {\bibfnamefont {D.}~\bibnamefont {Kameda}}, \bibinfo {author} {\bibfnamefont
  {Y.}~\bibnamefont {Kawada}}, \bibinfo {author} {\bibfnamefont
  {T.}~\bibnamefont {Kubo}}, \bibinfo {author} {\bibfnamefont {K.}~\bibnamefont
  {Kusaka}}, \bibinfo {author} {\bibfnamefont {A.}~\bibnamefont {Mengoni}},
  \bibinfo {author} {\bibfnamefont {T.}~\bibnamefont {Motobayashi}}, \bibinfo
  {author} {\bibfnamefont {T.}~\bibnamefont {Ohnishi}}, \bibinfo {author}
  {\bibfnamefont {M.}~\bibnamefont {Ohtake}}, \bibinfo {author} {\bibfnamefont
  {N.~A.}\ \bibnamefont {Orr}}, \bibinfo {author} {\bibfnamefont
  {H.}~\bibnamefont {Otsu}}, \bibinfo {author} {\bibfnamefont {T.}~\bibnamefont
  {Otsuka}}, \bibinfo {author} {\bibfnamefont {A.}~\bibnamefont {Saito}},
  \bibinfo {author} {\bibfnamefont {H.}~\bibnamefont {Sakurai}}, \bibinfo
  {author} {\bibfnamefont {S.}~\bibnamefont {Shimoura}}, \bibinfo {author}
  {\bibfnamefont {T.}~\bibnamefont {Sumikama}}, \bibinfo {author}
  {\bibfnamefont {H.}~\bibnamefont {Takeda}}, \bibinfo {author} {\bibfnamefont
  {E.}~\bibnamefont {Takeshita}}, \bibinfo {author} {\bibfnamefont
  {M.}~\bibnamefont {Takechi}}, \bibinfo {author} {\bibfnamefont
  {S.}~\bibnamefont {Takeuchi}}, \bibinfo {author} {\bibfnamefont
  {K.}~\bibnamefont {Tanaka}}, \bibinfo {author} {\bibfnamefont {K.~N.}\
  \bibnamefont {Tanaka}}, \bibinfo {author} {\bibfnamefont {N.}~\bibnamefont
  {Tanaka}}, \bibinfo {author} {\bibfnamefont {Y.}~\bibnamefont {Togano}},
  \bibinfo {author} {\bibfnamefont {Y.}~\bibnamefont {Utsuno}}, \bibinfo
  {author} {\bibfnamefont {K.}~\bibnamefont {Yoneda}}, \bibinfo {author}
  {\bibfnamefont {A.}~\bibnamefont {Yoshida}},\ and\ \bibinfo {author}
  {\bibfnamefont {K.}~\bibnamefont {Yoshida}},\ }\href
  {https://doi.org/10.1103/PhysRevLett.103.262501} {\bibfield  {journal}
  {\bibinfo  {journal} {Phys. Rev. Lett.}\ }\textbf {\bibinfo {volume} {103}},\
  \bibinfo {pages} {262501} (\bibinfo {year} {2009})}\BibitemShut {NoStop}%
\bibitem [{\citenamefont {Kobayashi}\ \emph {et~al.}(2014)\citenamefont
  {Kobayashi}, \citenamefont {Nakamura}, \citenamefont {Kondo}, \citenamefont
  {Tostevin}, \citenamefont {Utsuno}, \citenamefont {Aoi}, \citenamefont
  {Baba}, \citenamefont {Barthelemy}, \citenamefont {Famiano}, \citenamefont
  {Fukuda}, \citenamefont {Inabe}, \citenamefont {Ishihara}, \citenamefont
  {Kanungo}, \citenamefont {Kim}, \citenamefont {Kubo}, \citenamefont {Lee},
  \citenamefont {Lee}, \citenamefont {Matsushita}, \citenamefont {Motobayashi},
  \citenamefont {Ohnishi}, \citenamefont {Orr}, \citenamefont {Otsu},
  \citenamefont {Otsuka}, \citenamefont {Sako}, \citenamefont {Sakurai},
  \citenamefont {Satou}, \citenamefont {Sumikama}, \citenamefont {Takeda},
  \citenamefont {Takeuchi}, \citenamefont {Tanaka}, \citenamefont {Togano},\
  and\ \citenamefont {Yoneda}}]{Kobayashi2014PRL_242501}%
  \BibitemOpen
  \bibfield  {author} {\bibinfo {author} {\bibfnamefont {N.}~\bibnamefont
  {Kobayashi}}, \bibinfo {author} {\bibfnamefont {T.}~\bibnamefont {Nakamura}},
  \bibinfo {author} {\bibfnamefont {Y.}~\bibnamefont {Kondo}}, \bibinfo
  {author} {\bibfnamefont {J.~A.}\ \bibnamefont {Tostevin}}, \bibinfo {author}
  {\bibfnamefont {Y.}~\bibnamefont {Utsuno}}, \bibinfo {author} {\bibfnamefont
  {N.}~\bibnamefont {Aoi}}, \bibinfo {author} {\bibfnamefont {H.}~\bibnamefont
  {Baba}}, \bibinfo {author} {\bibfnamefont {R.}~\bibnamefont {Barthelemy}},
  \bibinfo {author} {\bibfnamefont {M.~A.}\ \bibnamefont {Famiano}}, \bibinfo
  {author} {\bibfnamefont {N.}~\bibnamefont {Fukuda}}, \bibinfo {author}
  {\bibfnamefont {N.}~\bibnamefont {Inabe}}, \bibinfo {author} {\bibfnamefont
  {M.}~\bibnamefont {Ishihara}}, \bibinfo {author} {\bibfnamefont
  {R.}~\bibnamefont {Kanungo}}, \bibinfo {author} {\bibfnamefont
  {S.}~\bibnamefont {Kim}}, \bibinfo {author} {\bibfnamefont {T.}~\bibnamefont
  {Kubo}}, \bibinfo {author} {\bibfnamefont {G.~S.}\ \bibnamefont {Lee}},
  \bibinfo {author} {\bibfnamefont {H.~S.}\ \bibnamefont {Lee}}, \bibinfo
  {author} {\bibfnamefont {M.}~\bibnamefont {Matsushita}}, \bibinfo {author}
  {\bibfnamefont {T.}~\bibnamefont {Motobayashi}}, \bibinfo {author}
  {\bibfnamefont {T.}~\bibnamefont {Ohnishi}}, \bibinfo {author} {\bibfnamefont
  {N.~A.}\ \bibnamefont {Orr}}, \bibinfo {author} {\bibfnamefont
  {H.}~\bibnamefont {Otsu}}, \bibinfo {author} {\bibfnamefont {T.}~\bibnamefont
  {Otsuka}}, \bibinfo {author} {\bibfnamefont {T.}~\bibnamefont {Sako}},
  \bibinfo {author} {\bibfnamefont {H.}~\bibnamefont {Sakurai}}, \bibinfo
  {author} {\bibfnamefont {Y.}~\bibnamefont {Satou}}, \bibinfo {author}
  {\bibfnamefont {T.}~\bibnamefont {Sumikama}}, \bibinfo {author}
  {\bibfnamefont {H.}~\bibnamefont {Takeda}}, \bibinfo {author} {\bibfnamefont
  {S.}~\bibnamefont {Takeuchi}}, \bibinfo {author} {\bibfnamefont
  {R.}~\bibnamefont {Tanaka}}, \bibinfo {author} {\bibfnamefont
  {Y.}~\bibnamefont {Togano}},\ and\ \bibinfo {author} {\bibfnamefont
  {K.}~\bibnamefont {Yoneda}},\ }\href
  {https://doi.org/10.1103/PhysRevLett.112.242501} {\bibfield  {journal}
  {\bibinfo  {journal} {Phys. Rev. Lett.}\ }\textbf {\bibinfo {volume} {112}},\
  \bibinfo {pages} {242501} (\bibinfo {year} {2014})}\BibitemShut {NoStop}%
\bibitem [{\citenamefont {Tostevin}\ and\ \citenamefont
  {Gade}(2014)}]{Tostevin2014PRC}%
  \BibitemOpen
  \bibfield  {author} {\bibinfo {author} {\bibfnamefont {J.~A.}\ \bibnamefont
  {Tostevin}}\ and\ \bibinfo {author} {\bibfnamefont {A.}~\bibnamefont
  {Gade}},\ }\href {https://doi.org/10.1103/PhysRevC.90.057602} {\bibfield
  {journal} {\bibinfo  {journal} {Phys. Rev. C}\ }\textbf {\bibinfo {volume}
  {90}},\ \bibinfo {pages} {057602} (\bibinfo {year} {2014})}\BibitemShut
  {NoStop}%
\bibitem [{\citenamefont {Tostevin}\ and\ \citenamefont
  {Gade}(2021)}]{Tostevin2021PRC}%
  \BibitemOpen
  \bibfield  {author} {\bibinfo {author} {\bibfnamefont {J.~A.}\ \bibnamefont
  {Tostevin}}\ and\ \bibinfo {author} {\bibfnamefont {A.}~\bibnamefont
  {Gade}},\ }\href {https://doi.org/10.1103/PhysRevC.103.054610} {\bibfield
  {journal} {\bibinfo  {journal} {Phys. Rev. C}\ }\textbf {\bibinfo {volume}
  {103}},\ \bibinfo {pages} {054610} (\bibinfo {year} {2021})}\BibitemShut
  {NoStop}%
\bibitem [{\citenamefont {Tanihata}\ \emph {et~al.}(1985)\citenamefont
  {Tanihata}, \citenamefont {Hamagaki}, \citenamefont {Hashimoto},
  \citenamefont {Shida}, \citenamefont {Yoshikawa}, \citenamefont {Sugimoto},
  \citenamefont {Yamakawa}, \citenamefont {Kobayashi},\ and\ \citenamefont
  {Takahashi}}]{Tanihata1985PRL}%
  \BibitemOpen
  \bibfield  {author} {\bibinfo {author} {\bibfnamefont {I.}~\bibnamefont
  {Tanihata}}, \bibinfo {author} {\bibfnamefont {H.}~\bibnamefont {Hamagaki}},
  \bibinfo {author} {\bibfnamefont {O.}~\bibnamefont {Hashimoto}}, \bibinfo
  {author} {\bibfnamefont {Y.}~\bibnamefont {Shida}}, \bibinfo {author}
  {\bibfnamefont {N.}~\bibnamefont {Yoshikawa}}, \bibinfo {author}
  {\bibfnamefont {K.}~\bibnamefont {Sugimoto}}, \bibinfo {author}
  {\bibfnamefont {O.}~\bibnamefont {Yamakawa}}, \bibinfo {author}
  {\bibfnamefont {T.}~\bibnamefont {Kobayashi}},\ and\ \bibinfo {author}
  {\bibfnamefont {N.}~\bibnamefont {Takahashi}},\ }\href
  {https://doi.org/10.1103/PhysRevLett.55.2676} {\bibfield  {journal} {\bibinfo
   {journal} {Phys. Rev. Lett.}\ }\textbf {\bibinfo {volume} {55}},\ \bibinfo
  {pages} {2676} (\bibinfo {year} {1985})}\BibitemShut {NoStop}%
\bibitem [{\citenamefont {Fukuda}\ \emph {et~al.}(1991)\citenamefont {Fukuda},
  \citenamefont {Ichihara}, \citenamefont {Inabe}, \citenamefont {Kubo},
  \citenamefont {Kumagai}, \citenamefont {Nakagawa}, \citenamefont {Yano},
  \citenamefont {Tanihata}, \citenamefont {Adachi}, \citenamefont {Asahi},
  \citenamefont {Kouguchi}, \citenamefont {Ishihara}, \citenamefont {Sagawa},\
  and\ \citenamefont {Shimoura}}]{Fukuda1991PLB}%
  \BibitemOpen
  \bibfield  {author} {\bibinfo {author} {\bibfnamefont {M.}~\bibnamefont
  {Fukuda}}, \bibinfo {author} {\bibfnamefont {T.}~\bibnamefont {Ichihara}},
  \bibinfo {author} {\bibfnamefont {N.}~\bibnamefont {Inabe}}, \bibinfo
  {author} {\bibfnamefont {T.}~\bibnamefont {Kubo}}, \bibinfo {author}
  {\bibfnamefont {H.}~\bibnamefont {Kumagai}}, \bibinfo {author} {\bibfnamefont
  {T.}~\bibnamefont {Nakagawa}}, \bibinfo {author} {\bibfnamefont
  {Y.}~\bibnamefont {Yano}}, \bibinfo {author} {\bibfnamefont {I.}~\bibnamefont
  {Tanihata}}, \bibinfo {author} {\bibfnamefont {M.}~\bibnamefont {Adachi}},
  \bibinfo {author} {\bibfnamefont {K.}~\bibnamefont {Asahi}}, \bibinfo
  {author} {\bibfnamefont {M.}~\bibnamefont {Kouguchi}}, \bibinfo {author}
  {\bibfnamefont {M.}~\bibnamefont {Ishihara}}, \bibinfo {author}
  {\bibfnamefont {H.}~\bibnamefont {Sagawa}},\ and\ \bibinfo {author}
  {\bibfnamefont {S.}~\bibnamefont {Shimoura}},\ }\href
  {https://doi.org/10.1016/0370-2693(91)91587-L} {\bibfield  {journal}
  {\bibinfo  {journal} {Phys. Lett. B}\ }\textbf {\bibinfo {volume} {268}},\
  \bibinfo {pages} {339} (\bibinfo {year} {1991})}\BibitemShut {NoStop}%
\bibitem [{\citenamefont {Tanihata}\ \emph {et~al.}(1988)\citenamefont
  {Tanihata}, \citenamefont {Kobayashi}, \citenamefont {Yamakawa},
  \citenamefont {Shimoura}, \citenamefont {Ekuni}, \citenamefont {Sugimoto},
  \citenamefont {Takahashi}, \citenamefont {Shimoda},\ and\ \citenamefont
  {Sato}}]{Tanihata1988PLB}%
  \BibitemOpen
  \bibfield  {author} {\bibinfo {author} {\bibfnamefont {I.}~\bibnamefont
  {Tanihata}}, \bibinfo {author} {\bibfnamefont {T.}~\bibnamefont {Kobayashi}},
  \bibinfo {author} {\bibfnamefont {O.}~\bibnamefont {Yamakawa}}, \bibinfo
  {author} {\bibfnamefont {S.}~\bibnamefont {Shimoura}}, \bibinfo {author}
  {\bibfnamefont {K.}~\bibnamefont {Ekuni}}, \bibinfo {author} {\bibfnamefont
  {K.}~\bibnamefont {Sugimoto}}, \bibinfo {author} {\bibfnamefont
  {N.}~\bibnamefont {Takahashi}}, \bibinfo {author} {\bibfnamefont
  {T.}~\bibnamefont {Shimoda}},\ and\ \bibinfo {author} {\bibfnamefont
  {H.}~\bibnamefont {Sato}},\ }\href
  {https://doi.org/https://doi.org/10.1016/0370-2693(88)90702-2} {\bibfield
  {journal} {\bibinfo  {journal} {Physics Letters B}\ }\textbf {\bibinfo
  {volume} {206}},\ \bibinfo {pages} {592} (\bibinfo {year}
  {1988})}\BibitemShut {NoStop}%
\bibitem [{\citenamefont {Nakamura}\ \emph {et~al.}(1994)\citenamefont
  {Nakamura}, \citenamefont {Shimoura}, \citenamefont {Kobayashi},
  \citenamefont {Teranishi}, \citenamefont {Abe}, \citenamefont {Aoi},
  \citenamefont {Doki}, \citenamefont {Fujimaki}, \citenamefont {Inabe},
  \citenamefont {Ishihara}, \citenamefont {Iwasa}, \citenamefont {Kubo},
  \citenamefont {Mengoni}, \citenamefont {Motobayashi}, \citenamefont {Ozawa},
  \citenamefont {Sakurai}, \citenamefont {Watanabe},\ and\ \citenamefont
  {Watanabe}}]{Nakamura1994PLB}%
  \BibitemOpen
  \bibfield  {author} {\bibinfo {author} {\bibfnamefont {T.}~\bibnamefont
  {Nakamura}}, \bibinfo {author} {\bibfnamefont {S.}~\bibnamefont {Shimoura}},
  \bibinfo {author} {\bibfnamefont {T.}~\bibnamefont {Kobayashi}}, \bibinfo
  {author} {\bibfnamefont {T.}~\bibnamefont {Teranishi}}, \bibinfo {author}
  {\bibfnamefont {K.}~\bibnamefont {Abe}}, \bibinfo {author} {\bibfnamefont
  {N.}~\bibnamefont {Aoi}}, \bibinfo {author} {\bibfnamefont {Y.}~\bibnamefont
  {Doki}}, \bibinfo {author} {\bibfnamefont {M.}~\bibnamefont {Fujimaki}},
  \bibinfo {author} {\bibfnamefont {N.}~\bibnamefont {Inabe}}, \bibinfo
  {author} {\bibfnamefont {M.}~\bibnamefont {Ishihara}}, \bibinfo {author}
  {\bibfnamefont {N.}~\bibnamefont {Iwasa}}, \bibinfo {author} {\bibfnamefont
  {T.}~\bibnamefont {Kubo}}, \bibinfo {author} {\bibfnamefont {A.}~\bibnamefont
  {Mengoni}}, \bibinfo {author} {\bibfnamefont {T.}~\bibnamefont
  {Motobayashi}}, \bibinfo {author} {\bibfnamefont {A.}~\bibnamefont {Ozawa}},
  \bibinfo {author} {\bibfnamefont {H.}~\bibnamefont {Sakurai}}, \bibinfo
  {author} {\bibfnamefont {Y.~X.}\ \bibnamefont {Watanabe}},\ and\ \bibinfo
  {author} {\bibfnamefont {Y.}~\bibnamefont {Watanabe}},\ }\href
  {https://doi.org/10.1016/0370-2693(94)91055-3} {\bibfield  {journal}
  {\bibinfo  {journal} {Phys. Lett. B}\ }\textbf {\bibinfo {volume} {331}},\
  \bibinfo {pages} {296} (\bibinfo {year} {1994})}\BibitemShut {NoStop}%
\bibitem [{\citenamefont {Fauerbach}\ \emph {et~al.}(1997)\citenamefont
  {Fauerbach}, \citenamefont {Chromik}, \citenamefont {Glasmacher},
  \citenamefont {Hansen}, \citenamefont {Ibbotson}, \citenamefont {Morrissey},
  \citenamefont {Scheit}, \citenamefont {Thompson}, \citenamefont {Tostevin},
  \citenamefont {Tucholski},\ and\ \citenamefont
  {Vinciquerra}}]{Fauerbach1997PRC}%
  \BibitemOpen
  \bibfield  {author} {\bibinfo {author} {\bibfnamefont {M.}~\bibnamefont
  {Fauerbach}}, \bibinfo {author} {\bibfnamefont {M.}~\bibnamefont {Chromik}},
  \bibinfo {author} {\bibfnamefont {T.}~\bibnamefont {Glasmacher}}, \bibinfo
  {author} {\bibfnamefont {P.~G.}\ \bibnamefont {Hansen}}, \bibinfo {author}
  {\bibfnamefont {R.}~\bibnamefont {Ibbotson}}, \bibinfo {author}
  {\bibfnamefont {D.~J.}\ \bibnamefont {Morrissey}}, \bibinfo {author}
  {\bibfnamefont {H.}~\bibnamefont {Scheit}}, \bibinfo {author} {\bibfnamefont
  {I.~J.}\ \bibnamefont {Thompson}}, \bibinfo {author} {\bibfnamefont {J.~A.}\
  \bibnamefont {Tostevin}}, \bibinfo {author} {\bibfnamefont {A.}~\bibnamefont
  {Tucholski}},\ and\ \bibinfo {author} {\bibfnamefont {A.}~\bibnamefont
  {Vinciquerra}},\ }\href {https://doi.org/10.1103/PhysRevC.56.R1} {\bibfield
  {journal} {\bibinfo  {journal} {Phys. Rev. C}\ }\textbf {\bibinfo {volume}
  {56}},\ \bibinfo {pages} {R1} (\bibinfo {year} {1997})}\BibitemShut {NoStop}%
\bibitem [{\citenamefont {Hansen}\ and\ \citenamefont
  {Jonson}(1987)}]{hansen1987neutron}%
  \BibitemOpen
  \bibfield  {author} {\bibinfo {author} {\bibfnamefont {P.}~\bibnamefont
  {Hansen}}\ and\ \bibinfo {author} {\bibfnamefont {B.}~\bibnamefont
  {Jonson}},\ }\href {https://doi.org/10.1209/0295-5075/4/4/005} {\bibfield
  {journal} {\bibinfo  {journal} {EPL (Europhysics Letters)}\ }\textbf
  {\bibinfo {volume} {4}},\ \bibinfo {pages} {409} (\bibinfo {year}
  {1987})}\BibitemShut {NoStop}%
\bibitem [{\citenamefont {Jensen}\ \emph {et~al.}(2004)\citenamefont {Jensen},
  \citenamefont {Riisager}, \citenamefont {Fedorov},\ and\ \citenamefont
  {Garrido}}]{jensen2004structure}%
  \BibitemOpen
  \bibfield  {author} {\bibinfo {author} {\bibfnamefont {A.}~\bibnamefont
  {Jensen}}, \bibinfo {author} {\bibfnamefont {K.}~\bibnamefont {Riisager}},
  \bibinfo {author} {\bibfnamefont {D.}~\bibnamefont {Fedorov}},\ and\ \bibinfo
  {author} {\bibfnamefont {E.}~\bibnamefont {Garrido}},\ }\href
  {https://doi.org/10.1103/RevModPhys.76.215} {\bibfield  {journal} {\bibinfo
  {journal} {Reviews of Modern Physics}\ }\textbf {\bibinfo {volume} {76}},\
  \bibinfo {pages} {215} (\bibinfo {year} {2004})}\BibitemShut {NoStop}%
\bibitem [{\citenamefont {Yang}\ \emph {et~al.}(2021)\citenamefont {Yang},
  \citenamefont {Kubota}, \citenamefont {Corsi}, \citenamefont {Yoshida},
  \citenamefont {Sun}, \citenamefont {Li}, \citenamefont {Kimura},
  \citenamefont {Michel}, \citenamefont {Ogata}, \citenamefont {Yuan},
  \citenamefont {Yuan}, \citenamefont {Authelet}, \citenamefont {Baba},
  \citenamefont {Caesar}, \citenamefont {Calvet}, \citenamefont {Delbart},
  \citenamefont {Dozono}, \citenamefont {Feng}, \citenamefont {Flavigny},
  \citenamefont {Gheller}, \citenamefont {Gibelin}, \citenamefont {Giganon},
  \citenamefont {Gillibert}, \citenamefont {Hasegawa}, \citenamefont {Isobe},
  \citenamefont {Kanaya}, \citenamefont {Kawakami}, \citenamefont {Kim},
  \citenamefont {Kiyokawa}, \citenamefont {Kobayashi}, \citenamefont
  {Kobayashi}, \citenamefont {Kobayashi}, \citenamefont {Kondo}, \citenamefont
  {Korkulu}, \citenamefont {Koyama}, \citenamefont {Lapoux}, \citenamefont
  {Maeda}, \citenamefont {Marqu\'es}, \citenamefont {Motobayashi},
  \citenamefont {Miyazaki}, \citenamefont {Nakamura}, \citenamefont
  {Nakatsuka}, \citenamefont {Nishio}, \citenamefont {Obertelli}, \citenamefont
  {Ohkura}, \citenamefont {Orr}, \citenamefont {Ota}, \citenamefont {Otsu},
  \citenamefont {Ozaki}, \citenamefont {Panin}, \citenamefont {Paschalis},
  \citenamefont {Pollacco}, \citenamefont {Reichert}, \citenamefont {Rouss\'e},
  \citenamefont {Saito}, \citenamefont {Sakaguchi}, \citenamefont {Sako},
  \citenamefont {Santamaria}, \citenamefont {Sasano}, \citenamefont {Sato},
  \citenamefont {Shikata}, \citenamefont {Shimizu}, \citenamefont {Shindo},
  \citenamefont {Stuhl}, \citenamefont {Sumikama}, \citenamefont {Sun},
  \citenamefont {Tabata}, \citenamefont {Togano}, \citenamefont {Tsubota},
  \citenamefont {Xu}, \citenamefont {Yasuda}, \citenamefont {Yoneda},
  \citenamefont {Zenihiro}, \citenamefont {Zhou}, \citenamefont {Zuo},\ and\
  \citenamefont {Uesaka}}]{Yang2021PRL}%
  \BibitemOpen
  \bibfield  {author} {\bibinfo {author} {\bibfnamefont {Z.~H.}\ \bibnamefont
  {Yang}}, \bibinfo {author} {\bibfnamefont {Y.}~\bibnamefont {Kubota}},
  \bibinfo {author} {\bibfnamefont {A.}~\bibnamefont {Corsi}}, \bibinfo
  {author} {\bibfnamefont {K.}~\bibnamefont {Yoshida}}, \bibinfo {author}
  {\bibfnamefont {X.-X.}\ \bibnamefont {Sun}}, \bibinfo {author} {\bibfnamefont
  {J.~G.}\ \bibnamefont {Li}}, \bibinfo {author} {\bibfnamefont
  {M.}~\bibnamefont {Kimura}}, \bibinfo {author} {\bibfnamefont
  {N.}~\bibnamefont {Michel}}, \bibinfo {author} {\bibfnamefont
  {K.}~\bibnamefont {Ogata}}, \bibinfo {author} {\bibfnamefont {C.~X.}\
  \bibnamefont {Yuan}}, \bibinfo {author} {\bibfnamefont {Q.}~\bibnamefont
  {Yuan}}, \bibinfo {author} {\bibfnamefont {G.}~\bibnamefont {Authelet}},
  \bibinfo {author} {\bibfnamefont {H.}~\bibnamefont {Baba}}, \bibinfo {author}
  {\bibfnamefont {C.}~\bibnamefont {Caesar}}, \bibinfo {author} {\bibfnamefont
  {D.}~\bibnamefont {Calvet}}, \bibinfo {author} {\bibfnamefont
  {A.}~\bibnamefont {Delbart}}, \bibinfo {author} {\bibfnamefont
  {M.}~\bibnamefont {Dozono}}, \bibinfo {author} {\bibfnamefont
  {J.}~\bibnamefont {Feng}}, \bibinfo {author} {\bibfnamefont {F.}~\bibnamefont
  {Flavigny}}, \bibinfo {author} {\bibfnamefont {J.-M.}\ \bibnamefont
  {Gheller}}, \bibinfo {author} {\bibfnamefont {J.}~\bibnamefont {Gibelin}},
  \bibinfo {author} {\bibfnamefont {A.}~\bibnamefont {Giganon}}, \bibinfo
  {author} {\bibfnamefont {A.}~\bibnamefont {Gillibert}}, \bibinfo {author}
  {\bibfnamefont {K.}~\bibnamefont {Hasegawa}}, \bibinfo {author}
  {\bibfnamefont {T.}~\bibnamefont {Isobe}}, \bibinfo {author} {\bibfnamefont
  {Y.}~\bibnamefont {Kanaya}}, \bibinfo {author} {\bibfnamefont
  {S.}~\bibnamefont {Kawakami}}, \bibinfo {author} {\bibfnamefont
  {D.}~\bibnamefont {Kim}}, \bibinfo {author} {\bibfnamefont {Y.}~\bibnamefont
  {Kiyokawa}}, \bibinfo {author} {\bibfnamefont {M.}~\bibnamefont {Kobayashi}},
  \bibinfo {author} {\bibfnamefont {N.}~\bibnamefont {Kobayashi}}, \bibinfo
  {author} {\bibfnamefont {T.}~\bibnamefont {Kobayashi}}, \bibinfo {author}
  {\bibfnamefont {Y.}~\bibnamefont {Kondo}}, \bibinfo {author} {\bibfnamefont
  {Z.}~\bibnamefont {Korkulu}}, \bibinfo {author} {\bibfnamefont
  {S.}~\bibnamefont {Koyama}}, \bibinfo {author} {\bibfnamefont
  {V.}~\bibnamefont {Lapoux}}, \bibinfo {author} {\bibfnamefont
  {Y.}~\bibnamefont {Maeda}}, \bibinfo {author} {\bibfnamefont {F.~M.}\
  \bibnamefont {Marqu\'es}}, \bibinfo {author} {\bibfnamefont {T.}~\bibnamefont
  {Motobayashi}}, \bibinfo {author} {\bibfnamefont {T.}~\bibnamefont
  {Miyazaki}}, \bibinfo {author} {\bibfnamefont {T.}~\bibnamefont {Nakamura}},
  \bibinfo {author} {\bibfnamefont {N.}~\bibnamefont {Nakatsuka}}, \bibinfo
  {author} {\bibfnamefont {Y.}~\bibnamefont {Nishio}}, \bibinfo {author}
  {\bibfnamefont {A.}~\bibnamefont {Obertelli}}, \bibinfo {author}
  {\bibfnamefont {A.}~\bibnamefont {Ohkura}}, \bibinfo {author} {\bibfnamefont
  {N.~A.}\ \bibnamefont {Orr}}, \bibinfo {author} {\bibfnamefont
  {S.}~\bibnamefont {Ota}}, \bibinfo {author} {\bibfnamefont {H.}~\bibnamefont
  {Otsu}}, \bibinfo {author} {\bibfnamefont {T.}~\bibnamefont {Ozaki}},
  \bibinfo {author} {\bibfnamefont {V.}~\bibnamefont {Panin}}, \bibinfo
  {author} {\bibfnamefont {S.}~\bibnamefont {Paschalis}}, \bibinfo {author}
  {\bibfnamefont {E.~C.}\ \bibnamefont {Pollacco}}, \bibinfo {author}
  {\bibfnamefont {S.}~\bibnamefont {Reichert}}, \bibinfo {author}
  {\bibfnamefont {J.-Y.}\ \bibnamefont {Rouss\'e}}, \bibinfo {author}
  {\bibfnamefont {A.~T.}\ \bibnamefont {Saito}}, \bibinfo {author}
  {\bibfnamefont {S.}~\bibnamefont {Sakaguchi}}, \bibinfo {author}
  {\bibfnamefont {M.}~\bibnamefont {Sako}}, \bibinfo {author} {\bibfnamefont
  {C.}~\bibnamefont {Santamaria}}, \bibinfo {author} {\bibfnamefont
  {M.}~\bibnamefont {Sasano}}, \bibinfo {author} {\bibfnamefont
  {H.}~\bibnamefont {Sato}}, \bibinfo {author} {\bibfnamefont {M.}~\bibnamefont
  {Shikata}}, \bibinfo {author} {\bibfnamefont {Y.}~\bibnamefont {Shimizu}},
  \bibinfo {author} {\bibfnamefont {Y.}~\bibnamefont {Shindo}}, \bibinfo
  {author} {\bibfnamefont {L.}~\bibnamefont {Stuhl}}, \bibinfo {author}
  {\bibfnamefont {T.}~\bibnamefont {Sumikama}}, \bibinfo {author}
  {\bibfnamefont {Y.~L.}\ \bibnamefont {Sun}}, \bibinfo {author} {\bibfnamefont
  {M.}~\bibnamefont {Tabata}}, \bibinfo {author} {\bibfnamefont
  {Y.}~\bibnamefont {Togano}}, \bibinfo {author} {\bibfnamefont
  {J.}~\bibnamefont {Tsubota}}, \bibinfo {author} {\bibfnamefont {F.~R.}\
  \bibnamefont {Xu}}, \bibinfo {author} {\bibfnamefont {J.}~\bibnamefont
  {Yasuda}}, \bibinfo {author} {\bibfnamefont {K.}~\bibnamefont {Yoneda}},
  \bibinfo {author} {\bibfnamefont {J.}~\bibnamefont {Zenihiro}}, \bibinfo
  {author} {\bibfnamefont {S.-G.}\ \bibnamefont {Zhou}}, \bibinfo {author}
  {\bibfnamefont {W.}~\bibnamefont {Zuo}},\ and\ \bibinfo {author}
  {\bibfnamefont {T.}~\bibnamefont {Uesaka}},\ }\href
  {https://doi.org/10.1103/PhysRevLett.126.082501} {\bibfield  {journal}
  {\bibinfo  {journal} {Phys. Rev. Lett.}\ }\textbf {\bibinfo {volume} {126}},\
  \bibinfo {pages} {082501} (\bibinfo {year} {2021})}\BibitemShut {NoStop}%
\bibitem [{\citenamefont {Freer}\ \emph {et~al.}(2018)\citenamefont {Freer},
  \citenamefont {Horiuchi}, \citenamefont {{Kanada-En'yo}}, \citenamefont
  {Lee},\ and\ \citenamefont {Mei{\ss}ner}}]{Freer2018RMP}%
  \BibitemOpen
  \bibfield  {author} {\bibinfo {author} {\bibfnamefont {M.}~\bibnamefont
  {Freer}}, \bibinfo {author} {\bibfnamefont {H.}~\bibnamefont {Horiuchi}},
  \bibinfo {author} {\bibfnamefont {Y.}~\bibnamefont {{Kanada-En'yo}}},
  \bibinfo {author} {\bibfnamefont {D.}~\bibnamefont {Lee}},\ and\ \bibinfo
  {author} {\bibfnamefont {U.-G.}\ \bibnamefont {Mei{\ss}ner}},\ }\href
  {https://doi.org/10.1103/RevModPhys.90.035004} {\bibfield  {journal}
  {\bibinfo  {journal} {Rev. Mod. Phys.}\ }\textbf {\bibinfo {volume} {90}},\
  \bibinfo {pages} {035004} (\bibinfo {year} {2018})}\BibitemShut {NoStop}%
\bibitem [{\citenamefont {Shen}\ \emph {et~al.}(2025)\citenamefont {Shen},
  \citenamefont {Elhatisari}, \citenamefont {Lee}, \citenamefont
  {Mei{\ss}ner},\ and\ \citenamefont {Ren}}]{Shen:2024qzi}%
  \BibitemOpen
  \bibfield  {author} {\bibinfo {author} {\bibfnamefont {S.}~\bibnamefont
  {Shen}}, \bibinfo {author} {\bibfnamefont {S.}~\bibnamefont {Elhatisari}},
  \bibinfo {author} {\bibfnamefont {D.}~\bibnamefont {Lee}}, \bibinfo {author}
  {\bibfnamefont {U.-G.}\ \bibnamefont {Mei{\ss}ner}},\ and\ \bibinfo {author}
  {\bibfnamefont {Z.}~\bibnamefont {Ren}},\ }\href
  {https://doi.org/10.1103/PhysRevLett.134.162503} {\bibfield  {journal}
  {\bibinfo  {journal} {Phys. Rev. Lett.}\ }\textbf {\bibinfo {volume} {134}},\
  \bibinfo {pages} {162503} (\bibinfo {year} {2025})},\ \Eprint
  {https://arxiv.org/abs/2411.14935} {arXiv:2411.14935 [nucl-th]} \BibitemShut
  {NoStop}%
\bibitem [{\citenamefont {Takechi}\ \emph {et~al.}(2014)\citenamefont
  {Takechi}, \citenamefont {Suzuki}, \citenamefont {Nishimura}, \citenamefont
  {Fukuda}, \citenamefont {Ohtsubo}, \citenamefont {Nagashima}, \citenamefont
  {Suzuki}, \citenamefont {Yamaguchi}, \citenamefont {Ozawa}, \citenamefont
  {Moriguchi}, \citenamefont {Ohishi}, \citenamefont {Sumikama}, \citenamefont
  {Geissel}, \citenamefont {Aoi}, \citenamefont {Chen}, \citenamefont {Fang},
  \citenamefont {Fukuda}, \citenamefont {Fukuoka}, \citenamefont {Furuki},
  \citenamefont {Inabe}, \citenamefont {Ishibashi}, \citenamefont {Itoh},
  \citenamefont {Izumikawa}, \citenamefont {Kameda}, \citenamefont {Kubo},
  \citenamefont {Lantz}, \citenamefont {Lee}, \citenamefont {Ma}, \citenamefont
  {Matsuta}, \citenamefont {Mihara}, \citenamefont {Momota}, \citenamefont
  {Nagae}, \citenamefont {Nishikiori}, \citenamefont {Niwa}, \citenamefont
  {Ohnishi}, \citenamefont {Okumura}, \citenamefont {Ohtake}, \citenamefont
  {Ogura}, \citenamefont {Sakurai}, \citenamefont {Sato}, \citenamefont
  {Shimbara}, \citenamefont {Suzuki}, \citenamefont {Takeda}, \citenamefont
  {Takeuchi}, \citenamefont {Tanaka}, \citenamefont {Tanaka}, \citenamefont
  {Uenishi}, \citenamefont {Winkler}, \citenamefont {Yanagisawa}, \citenamefont
  {Watanabe}, \citenamefont {Minomo}, \citenamefont {Tagami}, \citenamefont
  {Shimada}, \citenamefont {Kimura}, \citenamefont {Matsumoto}, \citenamefont
  {Shimizu},\ and\ \citenamefont {Yahiro}}]{Takechi2014PRC_061305}%
  \BibitemOpen
  \bibfield  {author} {\bibinfo {author} {\bibfnamefont {M.}~\bibnamefont
  {Takechi}}, \bibinfo {author} {\bibfnamefont {S.}~\bibnamefont {Suzuki}},
  \bibinfo {author} {\bibfnamefont {D.}~\bibnamefont {Nishimura}}, \bibinfo
  {author} {\bibfnamefont {M.}~\bibnamefont {Fukuda}}, \bibinfo {author}
  {\bibfnamefont {T.}~\bibnamefont {Ohtsubo}}, \bibinfo {author} {\bibfnamefont
  {M.}~\bibnamefont {Nagashima}}, \bibinfo {author} {\bibfnamefont
  {T.}~\bibnamefont {Suzuki}}, \bibinfo {author} {\bibfnamefont
  {T.}~\bibnamefont {Yamaguchi}}, \bibinfo {author} {\bibfnamefont
  {A.}~\bibnamefont {Ozawa}}, \bibinfo {author} {\bibfnamefont
  {T.}~\bibnamefont {Moriguchi}}, \bibinfo {author} {\bibfnamefont
  {H.}~\bibnamefont {Ohishi}}, \bibinfo {author} {\bibfnamefont
  {T.}~\bibnamefont {Sumikama}}, \bibinfo {author} {\bibfnamefont
  {H.}~\bibnamefont {Geissel}}, \bibinfo {author} {\bibfnamefont
  {N.}~\bibnamefont {Aoi}}, \bibinfo {author} {\bibfnamefont {R.-J.}\
  \bibnamefont {Chen}}, \bibinfo {author} {\bibfnamefont {D.-Q.}\ \bibnamefont
  {Fang}}, \bibinfo {author} {\bibfnamefont {N.}~\bibnamefont {Fukuda}},
  \bibinfo {author} {\bibfnamefont {S.}~\bibnamefont {Fukuoka}}, \bibinfo
  {author} {\bibfnamefont {H.}~\bibnamefont {Furuki}}, \bibinfo {author}
  {\bibfnamefont {N.}~\bibnamefont {Inabe}}, \bibinfo {author} {\bibfnamefont
  {Y.}~\bibnamefont {Ishibashi}}, \bibinfo {author} {\bibfnamefont
  {T.}~\bibnamefont {Itoh}}, \bibinfo {author} {\bibfnamefont {T.}~\bibnamefont
  {Izumikawa}}, \bibinfo {author} {\bibfnamefont {D.}~\bibnamefont {Kameda}},
  \bibinfo {author} {\bibfnamefont {T.}~\bibnamefont {Kubo}}, \bibinfo {author}
  {\bibfnamefont {M.}~\bibnamefont {Lantz}}, \bibinfo {author} {\bibfnamefont
  {C.~S.}\ \bibnamefont {Lee}}, \bibinfo {author} {\bibfnamefont {Y.-G.}\
  \bibnamefont {Ma}}, \bibinfo {author} {\bibfnamefont {K.}~\bibnamefont
  {Matsuta}}, \bibinfo {author} {\bibfnamefont {M.}~\bibnamefont {Mihara}},
  \bibinfo {author} {\bibfnamefont {S.}~\bibnamefont {Momota}}, \bibinfo
  {author} {\bibfnamefont {D.}~\bibnamefont {Nagae}}, \bibinfo {author}
  {\bibfnamefont {R.}~\bibnamefont {Nishikiori}}, \bibinfo {author}
  {\bibfnamefont {T.}~\bibnamefont {Niwa}}, \bibinfo {author} {\bibfnamefont
  {T.}~\bibnamefont {Ohnishi}}, \bibinfo {author} {\bibfnamefont
  {K.}~\bibnamefont {Okumura}}, \bibinfo {author} {\bibfnamefont
  {M.}~\bibnamefont {Ohtake}}, \bibinfo {author} {\bibfnamefont
  {T.}~\bibnamefont {Ogura}}, \bibinfo {author} {\bibfnamefont
  {H.}~\bibnamefont {Sakurai}}, \bibinfo {author} {\bibfnamefont
  {K.}~\bibnamefont {Sato}}, \bibinfo {author} {\bibfnamefont {Y.}~\bibnamefont
  {Shimbara}}, \bibinfo {author} {\bibfnamefont {H.}~\bibnamefont {Suzuki}},
  \bibinfo {author} {\bibfnamefont {H.}~\bibnamefont {Takeda}}, \bibinfo
  {author} {\bibfnamefont {S.}~\bibnamefont {Takeuchi}}, \bibinfo {author}
  {\bibfnamefont {K.}~\bibnamefont {Tanaka}}, \bibinfo {author} {\bibfnamefont
  {M.}~\bibnamefont {Tanaka}}, \bibinfo {author} {\bibfnamefont
  {H.}~\bibnamefont {Uenishi}}, \bibinfo {author} {\bibfnamefont
  {M.}~\bibnamefont {Winkler}}, \bibinfo {author} {\bibfnamefont
  {Y.}~\bibnamefont {Yanagisawa}}, \bibinfo {author} {\bibfnamefont
  {S.}~\bibnamefont {Watanabe}}, \bibinfo {author} {\bibfnamefont
  {K.}~\bibnamefont {Minomo}}, \bibinfo {author} {\bibfnamefont
  {S.}~\bibnamefont {Tagami}}, \bibinfo {author} {\bibfnamefont
  {M.}~\bibnamefont {Shimada}}, \bibinfo {author} {\bibfnamefont
  {M.}~\bibnamefont {Kimura}}, \bibinfo {author} {\bibfnamefont
  {T.}~\bibnamefont {Matsumoto}}, \bibinfo {author} {\bibfnamefont {Y.~R.}\
  \bibnamefont {Shimizu}},\ and\ \bibinfo {author} {\bibfnamefont
  {M.}~\bibnamefont {Yahiro}},\ }\href
  {https://doi.org/10.1103/PhysRevC.90.061305} {\bibfield  {journal} {\bibinfo
  {journal} {Phys. Rev. C}\ }\textbf {\bibinfo {volume} {90}},\ \bibinfo
  {pages} {061305} (\bibinfo {year} {2014})}\BibitemShut {NoStop}%
\bibitem [{\citenamefont {Zhou}\ \emph {et~al.}(2010)\citenamefont {Zhou},
  \citenamefont {Meng}, \citenamefont {Ring},\ and\ \citenamefont
  {Zhao}}]{Zhou2010PRC(R)}%
  \BibitemOpen
  \bibfield  {author} {\bibinfo {author} {\bibfnamefont {S.-G.}\ \bibnamefont
  {Zhou}}, \bibinfo {author} {\bibfnamefont {J.}~\bibnamefont {Meng}}, \bibinfo
  {author} {\bibfnamefont {P.}~\bibnamefont {Ring}},\ and\ \bibinfo {author}
  {\bibfnamefont {E.-G.}\ \bibnamefont {Zhao}},\ }\href
  {https://doi.org/10.1103/PhysRevC.82.011301} {\bibfield  {journal} {\bibinfo
  {journal} {Phys. Rev. C}\ }\textbf {\bibinfo {volume} {82}},\ \bibinfo
  {pages} {011301(R)} (\bibinfo {year} {2010})}\BibitemShut {NoStop}%
\bibitem [{\citenamefont {Li}\ \emph {et~al.}(2012)\citenamefont {Li},
  \citenamefont {Meng}, \citenamefont {Ring}, \citenamefont {Zhao},\ and\
  \citenamefont {Zhou}}]{Li2012PRC}%
  \BibitemOpen
  \bibfield  {author} {\bibinfo {author} {\bibfnamefont {L.}~\bibnamefont
  {Li}}, \bibinfo {author} {\bibfnamefont {J.}~\bibnamefont {Meng}}, \bibinfo
  {author} {\bibfnamefont {P.}~\bibnamefont {Ring}}, \bibinfo {author}
  {\bibfnamefont {E.-G.}\ \bibnamefont {Zhao}},\ and\ \bibinfo {author}
  {\bibfnamefont {S.-G.}\ \bibnamefont {Zhou}},\ }\href
  {https://doi.org/10.1103/PhysRevC.85.024312} {\bibfield  {journal} {\bibinfo
  {journal} {Phys. Rev. C}\ }\textbf {\bibinfo {volume} {85}},\ \bibinfo
  {pages} {024312} (\bibinfo {year} {2012})}\BibitemShut {NoStop}%
\bibitem [{\citenamefont {Zhang}\ \emph {et~al.}(2020)\citenamefont {Zhang},
  \citenamefont {Cheoun}, \citenamefont {Choi}, \citenamefont {Chong},
  \citenamefont {Dong}, \citenamefont {Geng}, \citenamefont {Ha}, \citenamefont
  {He}, \citenamefont {Heo}, \citenamefont {Ho}, \citenamefont {In},
  \citenamefont {Kim}, \citenamefont {Kim}, \citenamefont {Lee}, \citenamefont
  {Lee}, \citenamefont {Li}, \citenamefont {Luo}, \citenamefont {Meng},
  \citenamefont {Mun}, \citenamefont {Niu}, \citenamefont {Pan}, \citenamefont
  {Papakonstantinou}, \citenamefont {Shang}, \citenamefont {Shen},
  \citenamefont {Shen}, \citenamefont {Sun}, \citenamefont {Sun}, \citenamefont
  {Tam}, \citenamefont {Thaivayongnou}, \citenamefont {Wang}, \citenamefont
  {Wong}, \citenamefont {Xia}, \citenamefont {Yan}, \citenamefont {Yeung},
  \citenamefont {Yiu}, \citenamefont {Zhang}, \citenamefont {Zhang},\ and\
  \citenamefont {Zhou}}]{Zhang2020PRC}%
  \BibitemOpen
  \bibfield  {author} {\bibinfo {author} {\bibfnamefont {K.}~\bibnamefont
  {Zhang}}, \bibinfo {author} {\bibfnamefont {M.-K.}\ \bibnamefont {Cheoun}},
  \bibinfo {author} {\bibfnamefont {Y.-B.}\ \bibnamefont {Choi}}, \bibinfo
  {author} {\bibfnamefont {P.~S.}\ \bibnamefont {Chong}}, \bibinfo {author}
  {\bibfnamefont {J.}~\bibnamefont {Dong}}, \bibinfo {author} {\bibfnamefont
  {L.}~\bibnamefont {Geng}}, \bibinfo {author} {\bibfnamefont {E.}~\bibnamefont
  {Ha}}, \bibinfo {author} {\bibfnamefont {X.}~\bibnamefont {He}}, \bibinfo
  {author} {\bibfnamefont {C.}~\bibnamefont {Heo}}, \bibinfo {author}
  {\bibfnamefont {M.~C.}\ \bibnamefont {Ho}}, \bibinfo {author} {\bibfnamefont
  {E.~J.}\ \bibnamefont {In}}, \bibinfo {author} {\bibfnamefont
  {S.}~\bibnamefont {Kim}}, \bibinfo {author} {\bibfnamefont {Y.}~\bibnamefont
  {Kim}}, \bibinfo {author} {\bibfnamefont {C.-H.}\ \bibnamefont {Lee}},
  \bibinfo {author} {\bibfnamefont {J.}~\bibnamefont {Lee}}, \bibinfo {author}
  {\bibfnamefont {Z.}~\bibnamefont {Li}}, \bibinfo {author} {\bibfnamefont
  {T.}~\bibnamefont {Luo}}, \bibinfo {author} {\bibfnamefont {J.}~\bibnamefont
  {Meng}}, \bibinfo {author} {\bibfnamefont {M.-H.}\ \bibnamefont {Mun}},
  \bibinfo {author} {\bibfnamefont {Z.}~\bibnamefont {Niu}}, \bibinfo {author}
  {\bibfnamefont {C.}~\bibnamefont {Pan}}, \bibinfo {author} {\bibfnamefont
  {P.}~\bibnamefont {Papakonstantinou}}, \bibinfo {author} {\bibfnamefont
  {X.}~\bibnamefont {Shang}}, \bibinfo {author} {\bibfnamefont
  {C.}~\bibnamefont {Shen}}, \bibinfo {author} {\bibfnamefont {G.}~\bibnamefont
  {Shen}}, \bibinfo {author} {\bibfnamefont {W.}~\bibnamefont {Sun}}, \bibinfo
  {author} {\bibfnamefont {X.-X.}\ \bibnamefont {Sun}}, \bibinfo {author}
  {\bibfnamefont {C.~K.}\ \bibnamefont {Tam}}, \bibinfo {author} {\bibnamefont
  {Thaivayongnou}}, \bibinfo {author} {\bibfnamefont {C.}~\bibnamefont {Wang}},
  \bibinfo {author} {\bibfnamefont {S.~H.}\ \bibnamefont {Wong}}, \bibinfo
  {author} {\bibfnamefont {X.}~\bibnamefont {Xia}}, \bibinfo {author}
  {\bibfnamefont {Y.}~\bibnamefont {Yan}}, \bibinfo {author} {\bibfnamefont
  {R.~W.-Y.}\ \bibnamefont {Yeung}}, \bibinfo {author} {\bibfnamefont {T.~C.}\
  \bibnamefont {Yiu}}, \bibinfo {author} {\bibfnamefont {S.}~\bibnamefont
  {Zhang}}, \bibinfo {author} {\bibfnamefont {W.}~\bibnamefont {Zhang}},\ and\
  \bibinfo {author} {\bibfnamefont {S.-G.}\ \bibnamefont {Zhou}} (\bibinfo
  {collaboration} {DRHBc Mass Table Collaboration}),\ }\href
  {https://doi.org/10.1103/PhysRevC.102.024314} {\bibfield  {journal} {\bibinfo
   {journal} {Phys. Rev. C}\ }\textbf {\bibinfo {volume} {102}},\ \bibinfo
  {pages} {024314} (\bibinfo {year} {2020})}\BibitemShut {NoStop}%
\bibitem [{\citenamefont {Pan}\ \emph {et~al.}(2022)\citenamefont {Pan},
  \citenamefont {Cheoun}, \citenamefont {Choi}, \citenamefont {Dong},
  \citenamefont {Du}, \citenamefont {Fan}, \citenamefont {Gao}, \citenamefont
  {Geng}, \citenamefont {Ha}, \citenamefont {He}, \citenamefont {Huang},
  \citenamefont {Huang}, \citenamefont {Kim}, \citenamefont {Kim},
  \citenamefont {Lee}, \citenamefont {Lee}, \citenamefont {Li}, \citenamefont
  {Liu}, \citenamefont {Ma}, \citenamefont {Meng}, \citenamefont {Mun},
  \citenamefont {Niu}, \citenamefont {Papakonstantinou}, \citenamefont {Shang},
  \citenamefont {Shen}, \citenamefont {Shen}, \citenamefont {Sun},
  \citenamefont {Sun}, \citenamefont {Wu}, \citenamefont {Wu}, \citenamefont
  {Xia}, \citenamefont {Yan}, \citenamefont {Yiu}, \citenamefont {Zhang},
  \citenamefont {Zhang}, \citenamefont {Zhang}, \citenamefont {Zhang},
  \citenamefont {Zhao}, \citenamefont {Zheng},\ and\ \citenamefont
  {Zhou}}]{Pan2022PRC}%
  \BibitemOpen
  \bibfield  {author} {\bibinfo {author} {\bibfnamefont {C.}~\bibnamefont
  {Pan}}, \bibinfo {author} {\bibfnamefont {M.-K.}\ \bibnamefont {Cheoun}},
  \bibinfo {author} {\bibfnamefont {Y.-B.}\ \bibnamefont {Choi}}, \bibinfo
  {author} {\bibfnamefont {J.}~\bibnamefont {Dong}}, \bibinfo {author}
  {\bibfnamefont {X.}~\bibnamefont {Du}}, \bibinfo {author} {\bibfnamefont
  {X.-H.}\ \bibnamefont {Fan}}, \bibinfo {author} {\bibfnamefont
  {W.}~\bibnamefont {Gao}}, \bibinfo {author} {\bibfnamefont {L.}~\bibnamefont
  {Geng}}, \bibinfo {author} {\bibfnamefont {E.}~\bibnamefont {Ha}}, \bibinfo
  {author} {\bibfnamefont {X.-T.}\ \bibnamefont {He}}, \bibinfo {author}
  {\bibfnamefont {J.}~\bibnamefont {Huang}}, \bibinfo {author} {\bibfnamefont
  {K.}~\bibnamefont {Huang}}, \bibinfo {author} {\bibfnamefont
  {S.}~\bibnamefont {Kim}}, \bibinfo {author} {\bibfnamefont {Y.}~\bibnamefont
  {Kim}}, \bibinfo {author} {\bibfnamefont {C.-H.}\ \bibnamefont {Lee}},
  \bibinfo {author} {\bibfnamefont {J.}~\bibnamefont {Lee}}, \bibinfo {author}
  {\bibfnamefont {Z.}~\bibnamefont {Li}}, \bibinfo {author} {\bibfnamefont
  {Z.-R.}\ \bibnamefont {Liu}}, \bibinfo {author} {\bibfnamefont
  {Y.}~\bibnamefont {Ma}}, \bibinfo {author} {\bibfnamefont {J.}~\bibnamefont
  {Meng}}, \bibinfo {author} {\bibfnamefont {M.-H.}\ \bibnamefont {Mun}},
  \bibinfo {author} {\bibfnamefont {Z.}~\bibnamefont {Niu}}, \bibinfo {author}
  {\bibfnamefont {P.}~\bibnamefont {Papakonstantinou}}, \bibinfo {author}
  {\bibfnamefont {X.}~\bibnamefont {Shang}}, \bibinfo {author} {\bibfnamefont
  {C.}~\bibnamefont {Shen}}, \bibinfo {author} {\bibfnamefont {G.}~\bibnamefont
  {Shen}}, \bibinfo {author} {\bibfnamefont {W.}~\bibnamefont {Sun}}, \bibinfo
  {author} {\bibfnamefont {X.-X.}\ \bibnamefont {Sun}}, \bibinfo {author}
  {\bibfnamefont {J.}~\bibnamefont {Wu}}, \bibinfo {author} {\bibfnamefont
  {X.}~\bibnamefont {Wu}}, \bibinfo {author} {\bibfnamefont {X.}~\bibnamefont
  {Xia}}, \bibinfo {author} {\bibfnamefont {Y.}~\bibnamefont {Yan}}, \bibinfo
  {author} {\bibfnamefont {T.~C.}\ \bibnamefont {Yiu}}, \bibinfo {author}
  {\bibfnamefont {K.}~\bibnamefont {Zhang}}, \bibinfo {author} {\bibfnamefont
  {S.}~\bibnamefont {Zhang}}, \bibinfo {author} {\bibfnamefont
  {W.}~\bibnamefont {Zhang}}, \bibinfo {author} {\bibfnamefont
  {X.}~\bibnamefont {Zhang}}, \bibinfo {author} {\bibfnamefont
  {Q.}~\bibnamefont {Zhao}}, \bibinfo {author} {\bibfnamefont {R.}~\bibnamefont
  {Zheng}},\ and\ \bibinfo {author} {\bibfnamefont {S.-G.}\ \bibnamefont
  {Zhou}} (\bibinfo {collaboration} {DRHBc Mass Table Collaboration}),\ }\href
  {https://doi.org/10.1103/PhysRevC.106.014316} {\bibfield  {journal} {\bibinfo
   {journal} {Phys. Rev. C}\ }\textbf {\bibinfo {volume} {106}},\ \bibinfo
  {pages} {014316} (\bibinfo {year} {2022})}\BibitemShut {NoStop}%
\bibitem [{\citenamefont {An}\ \emph {et~al.}(2024)\citenamefont {An},
  \citenamefont {Zhang}, \citenamefont {Lu}, \citenamefont {Zhong},\ and\
  \citenamefont {Zhang}}]{An2024}%
  \BibitemOpen
  \bibfield  {author} {\bibinfo {author} {\bibfnamefont {J.-L.}\ \bibnamefont
  {An}}, \bibinfo {author} {\bibfnamefont {K.-Y.}\ \bibnamefont {Zhang}},
  \bibinfo {author} {\bibfnamefont {Q.}~\bibnamefont {Lu}}, \bibinfo {author}
  {\bibfnamefont {S.-Y.}\ \bibnamefont {Zhong}},\ and\ \bibinfo {author}
  {\bibfnamefont {S.-S.}\ \bibnamefont {Zhang}},\ }\href
  {https://doi.org/https://doi.org/10.1016/j.physletb.2023.138422} {\bibfield
  {journal} {\bibinfo  {journal} {Physics Letters B}\ }\textbf {\bibinfo
  {volume} {849}},\ \bibinfo {pages} {138422} (\bibinfo {year}
  {2024})}\BibitemShut {NoStop}%
\bibitem [{\citenamefont {Horiuchi}\ \emph
  {et~al.}(2026{\natexlab{a}})\citenamefont {Horiuchi}, \citenamefont
  {Suzuki},\ and\ \citenamefont {Wiringa}}]{Horiuchi2026PRL}%
  \BibitemOpen
  \bibfield  {author} {\bibinfo {author} {\bibfnamefont {W.}~\bibnamefont
  {Horiuchi}}, \bibinfo {author} {\bibfnamefont {Y.}~\bibnamefont {Suzuki}},\
  and\ \bibinfo {author} {\bibfnamefont {R.~B.}\ \bibnamefont {Wiringa}},\
  }\href {https://doi.org/10.1103/ppqx-yn59} {\bibfield  {journal} {\bibinfo
  {journal} {Phys. Rev. Lett.}\ }\textbf {\bibinfo {volume} {136}},\ \bibinfo
  {pages} {202501} (\bibinfo {year} {2026}{\natexlab{a}})}\BibitemShut
  {NoStop}%
\bibitem [{\citenamefont {Horiuchi}\ \emph
  {et~al.}(2026{\natexlab{b}})\citenamefont {Horiuchi}, \citenamefont
  {Suzuki},\ and\ \citenamefont {Wiringa}}]{Horiuchi2026PRC}%
  \BibitemOpen
  \bibfield  {author} {\bibinfo {author} {\bibfnamefont {W.}~\bibnamefont
  {Horiuchi}}, \bibinfo {author} {\bibfnamefont {Y.}~\bibnamefont {Suzuki}},\
  and\ \bibinfo {author} {\bibfnamefont {R.~B.}\ \bibnamefont {Wiringa}},\
  }\href {https://doi.org/10.1103/gcbk-s7tc} {\bibfield  {journal} {\bibinfo
  {journal} {Phys. Rev. C}\ }\textbf {\bibinfo {volume} {113}},\ \bibinfo
  {pages} {064601} (\bibinfo {year} {2026}{\natexlab{b}})},\ \Eprint
  {https://arxiv.org/abs/2512.20100} {arXiv:2512.20100 [nucl-th]} \BibitemShut
  {NoStop}%
\bibitem [{\citenamefont {Abdulhamid}\ \emph {et~al.}(2024)\citenamefont
  {Abdulhamid} \emph {et~al.}}]{STAR:2024wgy}%
  \BibitemOpen
  \bibfield  {author} {\bibinfo {author} {\bibfnamefont {M.~I.}\ \bibnamefont
  {Abdulhamid}} \emph {et~al.} (\bibinfo {collaboration} {STAR}),\ }\href
  {https://doi.org/10.1038/s41586-024-08097-2} {\bibfield  {journal} {\bibinfo
  {journal} {Nature}\ }\textbf {\bibinfo {volume} {635}},\ \bibinfo {pages}
  {67} (\bibinfo {year} {2024})},\ \Eprint {https://arxiv.org/abs/2401.06625}
  {arXiv:2401.06625 [nucl-ex]} \BibitemShut {NoStop}%
\bibitem [{\citenamefont {Giacalone}\ \emph {et~al.}(2025)\citenamefont
  {Giacalone} \emph {et~al.}}]{Giacalone:2024luz}%
  \BibitemOpen
  \bibfield  {author} {\bibinfo {author} {\bibfnamefont {G.}~\bibnamefont
  {Giacalone}} \emph {et~al.},\ }\href {https://doi.org/10.1103/k8rb-jgvq}
  {\bibfield  {journal} {\bibinfo  {journal} {Phys. Rev. Lett.}\ }\textbf
  {\bibinfo {volume} {135}},\ \bibinfo {pages} {012302} (\bibinfo {year}
  {2025})},\ \Eprint {https://arxiv.org/abs/2402.05995} {arXiv:2402.05995
  [nucl-th]} \BibitemShut {NoStop}%
\bibitem [{\citenamefont {Aad}\ \emph {et~al.}(2026)\citenamefont {Aad} \emph
  {et~al.}}]{ATLAS:2025nnt}%
  \BibitemOpen
  \bibfield  {author} {\bibinfo {author} {\bibfnamefont {G.}~\bibnamefont
  {Aad}} \emph {et~al.} (\bibinfo {collaboration} {ATLAS}),\ }\href
  {https://doi.org/10.1103/xqxz-8bhf} {\bibfield  {journal} {\bibinfo
  {journal} {Phys. Rev. C}\ }\textbf {\bibinfo {volume} {113}},\ \bibinfo
  {pages} {045205} (\bibinfo {year} {2026})},\ \Eprint
  {https://arxiv.org/abs/2509.05171} {arXiv:2509.05171 [nucl-ex]} \BibitemShut
  {NoStop}%
\bibitem [{\citenamefont {Abualrob}\ \emph {et~al.}(2025)\citenamefont
  {Abualrob} \emph {et~al.}}]{ALICE:2025luc}%
  \BibitemOpen
  \bibfield  {author} {\bibinfo {author} {\bibfnamefont {I.~J.}\ \bibnamefont
  {Abualrob}} \emph {et~al.} (\bibinfo {collaboration} {ALICE}),\ }\href@noop
  {} {\  (\bibinfo {year} {2025})},\ \Eprint {https://arxiv.org/abs/2509.06428}
  {arXiv:2509.06428 [nucl-ex]} \BibitemShut {NoStop}%
\bibitem [{\citenamefont {Shen}\ \emph {et~al.}(2026)\citenamefont {Shen},
  \citenamefont {Elhatisari}, \citenamefont {Lee}, \citenamefont
  {Mei{\ss}ner},\ and\ \citenamefont {Ren}}]{Shen:2026liw}%
  \BibitemOpen
  \bibfield  {author} {\bibinfo {author} {\bibfnamefont {S.}~\bibnamefont
  {Shen}}, \bibinfo {author} {\bibfnamefont {S.}~\bibnamefont {Elhatisari}},
  \bibinfo {author} {\bibfnamefont {D.}~\bibnamefont {Lee}}, \bibinfo {author}
  {\bibfnamefont {U.-G.}\ \bibnamefont {Mei{\ss}ner}},\ and\ \bibinfo {author}
  {\bibfnamefont {Z.}~\bibnamefont {Ren}},\ }\href
  {https://doi.org/10.3390/particles9010025} {\bibfield  {journal} {\bibinfo
  {journal} {Particles}\ }\textbf {\bibinfo {volume} {9}},\ \bibinfo {pages}
  {25} (\bibinfo {year} {2026})},\ \Eprint {https://arxiv.org/abs/2603.06978}
  {arXiv:2603.06978 [nucl-th]} \BibitemShut {NoStop}%
\bibitem [{\citenamefont {Christley}\ and\ \citenamefont
  {Tostevin}(1999)}]{Christley1999PRC}%
  \BibitemOpen
  \bibfield  {author} {\bibinfo {author} {\bibfnamefont {J.~A.}\ \bibnamefont
  {Christley}}\ and\ \bibinfo {author} {\bibfnamefont {J.~A.}\ \bibnamefont
  {Tostevin}},\ }\href {https://doi.org/10.1103/PhysRevC.59.2309} {\bibfield
  {journal} {\bibinfo  {journal} {Phys. Rev. C}\ }\textbf {\bibinfo {volume}
  {59}},\ \bibinfo {pages} {2309} (\bibinfo {year} {1999})}\BibitemShut
  {NoStop}%
\bibitem [{\citenamefont {Abu-Ibrahim}\ \emph {et~al.}(2003)\citenamefont
  {Abu-Ibrahim}, \citenamefont {Ogawa}, \citenamefont {Suzuki},\ and\
  \citenamefont {Tanihata}}]{abu2003cross}%
  \BibitemOpen
  \bibfield  {author} {\bibinfo {author} {\bibfnamefont {B.}~\bibnamefont
  {Abu-Ibrahim}}, \bibinfo {author} {\bibfnamefont {Y.}~\bibnamefont {Ogawa}},
  \bibinfo {author} {\bibfnamefont {Y.}~\bibnamefont {Suzuki}},\ and\ \bibinfo
  {author} {\bibfnamefont {I.}~\bibnamefont {Tanihata}},\ }\href
  {https://doi.org/10.1016/S0010-4655(02)00734-8} {\bibfield  {journal}
  {\bibinfo  {journal} {Computer Physics Communications}\ }\textbf {\bibinfo
  {volume} {151}},\ \bibinfo {pages} {369} (\bibinfo {year}
  {2003})}\BibitemShut {NoStop}%
\bibitem [{\citenamefont {Horiuchi}\ \emph {et~al.}(2012)\citenamefont
  {Horiuchi}, \citenamefont {Inakura}, \citenamefont {Nakatsukasa},\ and\
  \citenamefont {Suzuki}}]{Horiuchi2012PRC}%
  \BibitemOpen
  \bibfield  {author} {\bibinfo {author} {\bibfnamefont {W.}~\bibnamefont
  {Horiuchi}}, \bibinfo {author} {\bibfnamefont {T.}~\bibnamefont {Inakura}},
  \bibinfo {author} {\bibfnamefont {T.}~\bibnamefont {Nakatsukasa}},\ and\
  \bibinfo {author} {\bibfnamefont {Y.}~\bibnamefont {Suzuki}},\ }\href
  {https://doi.org/10.1103/PhysRevC.86.024614} {\bibfield  {journal} {\bibinfo
  {journal} {Phys. Rev. C}\ }\textbf {\bibinfo {volume} {86}},\ \bibinfo
  {pages} {024614} (\bibinfo {year} {2012})}\BibitemShut {NoStop}%
\bibitem [{\citenamefont {Urata}\ \emph {et~al.}(2012)\citenamefont {Urata},
  \citenamefont {Hagino},\ and\ \citenamefont {Sagawa}}]{Urata2012}%
  \BibitemOpen
  \bibfield  {author} {\bibinfo {author} {\bibfnamefont {Y.}~\bibnamefont
  {Urata}}, \bibinfo {author} {\bibfnamefont {K.}~\bibnamefont {Hagino}},\ and\
  \bibinfo {author} {\bibfnamefont {H.}~\bibnamefont {Sagawa}},\ }\href
  {https://doi.org/10.1103/PhysRevC.86.044613} {\bibfield  {journal} {\bibinfo
  {journal} {Phys. Rev. C}\ }\textbf {\bibinfo {volume} {86}},\ \bibinfo
  {pages} {044613} (\bibinfo {year} {2012})}\BibitemShut {NoStop}%
\bibitem [{\citenamefont {Simpson}\ and\ \citenamefont
  {Tostevin}(2012)}]{Simpson2012}%
  \BibitemOpen
  \bibfield  {author} {\bibinfo {author} {\bibfnamefont {E.~C.}\ \bibnamefont
  {Simpson}}\ and\ \bibinfo {author} {\bibfnamefont {J.~A.}\ \bibnamefont
  {Tostevin}},\ }\href {https://doi.org/10.1103/PhysRevC.86.054603} {\bibfield
  {journal} {\bibinfo  {journal} {Phys. Rev. C}\ }\textbf {\bibinfo {volume}
  {86}},\ \bibinfo {pages} {054603} (\bibinfo {year} {2012})}\BibitemShut
  {NoStop}%
\bibitem [{\citenamefont {Hong}\ \emph {et~al.}(2017)\citenamefont {Hong},
  \citenamefont {Bertulani},\ and\ \citenamefont {Kruppa}}]{Hong2017}%
  \BibitemOpen
  \bibfield  {author} {\bibinfo {author} {\bibfnamefont {J.}~\bibnamefont
  {Hong}}, \bibinfo {author} {\bibfnamefont {C.~A.}\ \bibnamefont
  {Bertulani}},\ and\ \bibinfo {author} {\bibfnamefont {A.~T.}\ \bibnamefont
  {Kruppa}},\ }\href {https://doi.org/10.1103/PhysRevC.96.064603} {\bibfield
  {journal} {\bibinfo  {journal} {Phys. Rev. C}\ }\textbf {\bibinfo {volume}
  {96}},\ \bibinfo {pages} {064603} (\bibinfo {year} {2017})}\BibitemShut
  {NoStop}%
\bibitem [{\citenamefont {Elhatisari}\ \emph {et~al.}(2024)\citenamefont
  {Elhatisari} \emph {et~al.}}]{Elhatisari:2022zrb}%
  \BibitemOpen
  \bibfield  {author} {\bibinfo {author} {\bibfnamefont {S.}~\bibnamefont
  {Elhatisari}} \emph {et~al.},\ }\href
  {https://doi.org/10.1038/s41586-024-07422-z} {\bibfield  {journal} {\bibinfo
  {journal} {Nature}\ }\textbf {\bibinfo {volume} {630}},\ \bibinfo {pages}
  {59} (\bibinfo {year} {2024})},\ \Eprint {https://arxiv.org/abs/2210.17488}
  {arXiv:2210.17488 [nucl-th]} \BibitemShut {NoStop}%
\bibitem [{\citenamefont {Elhatisari}\ \emph {et~al.}(2017)\citenamefont
  {Elhatisari}, \citenamefont {Epelbaum}, \citenamefont {Krebs}, \citenamefont
  {L{\"a}hde}, \citenamefont {Lee}, \citenamefont {Li}, \citenamefont {Lu},
  \citenamefont {Mei{\ss}ner},\ and\ \citenamefont
  {Rupak}}]{Elhatisari:2017eno}%
  \BibitemOpen
  \bibfield  {author} {\bibinfo {author} {\bibfnamefont {S.}~\bibnamefont
  {Elhatisari}}, \bibinfo {author} {\bibfnamefont {E.}~\bibnamefont
  {Epelbaum}}, \bibinfo {author} {\bibfnamefont {H.}~\bibnamefont {Krebs}},
  \bibinfo {author} {\bibfnamefont {T.~A.}\ \bibnamefont {L{\"a}hde}}, \bibinfo
  {author} {\bibfnamefont {D.}~\bibnamefont {Lee}}, \bibinfo {author}
  {\bibfnamefont {N.}~\bibnamefont {Li}}, \bibinfo {author} {\bibfnamefont
  {B.-n.}\ \bibnamefont {Lu}}, \bibinfo {author} {\bibfnamefont {U.-G.}\
  \bibnamefont {Mei{\ss}ner}},\ and\ \bibinfo {author} {\bibfnamefont
  {G.}~\bibnamefont {Rupak}},\ }\href
  {https://doi.org/10.1103/PhysRevLett.119.222505} {\bibfield  {journal}
  {\bibinfo  {journal} {Phys. Rev. Lett.}\ }\textbf {\bibinfo {volume} {119}},\
  \bibinfo {pages} {222505} (\bibinfo {year} {2017})},\ \Eprint
  {https://arxiv.org/abs/1702.05177} {arXiv:1702.05177 [nucl-th]} \BibitemShut
  {NoStop}%
\bibitem [{\citenamefont {Lu}\ \emph {et~al.}(2019)\citenamefont {Lu},
  \citenamefont {Li}, \citenamefont {Elhatisari}, \citenamefont {Lee},
  \citenamefont {Epelbaum},\ and\ \citenamefont {Mei{\ss}ner}}]{Lu:2018bat}%
  \BibitemOpen
  \bibfield  {author} {\bibinfo {author} {\bibfnamefont {B.-N.}\ \bibnamefont
  {Lu}}, \bibinfo {author} {\bibfnamefont {N.}~\bibnamefont {Li}}, \bibinfo
  {author} {\bibfnamefont {S.}~\bibnamefont {Elhatisari}}, \bibinfo {author}
  {\bibfnamefont {D.}~\bibnamefont {Lee}}, \bibinfo {author} {\bibfnamefont
  {E.}~\bibnamefont {Epelbaum}},\ and\ \bibinfo {author} {\bibfnamefont
  {U.-G.}\ \bibnamefont {Mei{\ss}ner}},\ }\href
  {https://doi.org/10.1016/j.physletb.2019.134863} {\bibfield  {journal}
  {\bibinfo  {journal} {Phys. Lett. B}\ }\textbf {\bibinfo {volume} {797}},\
  \bibinfo {pages} {134863} (\bibinfo {year} {2019})},\ \Eprint
  {https://arxiv.org/abs/1812.10928} {arXiv:1812.10928 [nucl-th]} \BibitemShut
  {NoStop}%
\bibitem [{\citenamefont {Glauber}(1959)}]{glauber1959lectures}%
  \BibitemOpen
  \bibfield  {author} {\bibinfo {author} {\bibfnamefont {R.}~\bibnamefont
  {Glauber}},\ }\href@noop {} {\emph {\bibinfo {title} {Lectures in theoretical
  physics}}}\ (\bibinfo  {publisher} {Interscience Publishers, New York},\
  \bibinfo {year} {1959})\BibitemShut {NoStop}%
\bibitem [{\citenamefont {Blanpied}\ \emph {et~al.}(1981)\citenamefont
  {Blanpied}, \citenamefont {Hoffmann}, \citenamefont {Barlett}, \citenamefont
  {McGill}, \citenamefont {Greene}, \citenamefont {Ray}, \citenamefont
  {Van~Dyck}, \citenamefont {Amann},\ and\ \citenamefont
  {Thiessen}}]{Blanpied1981PRC}%
  \BibitemOpen
  \bibfield  {author} {\bibinfo {author} {\bibfnamefont {G.~S.}\ \bibnamefont
  {Blanpied}}, \bibinfo {author} {\bibfnamefont {G.~W.}\ \bibnamefont
  {Hoffmann}}, \bibinfo {author} {\bibfnamefont {M.~L.}\ \bibnamefont
  {Barlett}}, \bibinfo {author} {\bibfnamefont {J.~A.}\ \bibnamefont {McGill}},
  \bibinfo {author} {\bibfnamefont {S.~J.}\ \bibnamefont {Greene}}, \bibinfo
  {author} {\bibfnamefont {L.}~\bibnamefont {Ray}}, \bibinfo {author}
  {\bibfnamefont {O.~B.}\ \bibnamefont {Van~Dyck}}, \bibinfo {author}
  {\bibfnamefont {J.}~\bibnamefont {Amann}},\ and\ \bibinfo {author}
  {\bibfnamefont {H.~A.}\ \bibnamefont {Thiessen}},\ }\href
  {https://doi.org/10.1103/PhysRevC.23.2599} {\bibfield  {journal} {\bibinfo
  {journal} {Phys. Rev. C}\ }\textbf {\bibinfo {volume} {23}},\ \bibinfo
  {pages} {2599} (\bibinfo {year} {1981})}\BibitemShut {NoStop}%
\bibitem [{\citenamefont {Kohama}\ \emph {et~al.}(2008)\citenamefont {Kohama},
  \citenamefont {Iida},\ and\ \citenamefont {Oyamatsu}}]{Kohama2008PRC}%
  \BibitemOpen
  \bibfield  {author} {\bibinfo {author} {\bibfnamefont {A.}~\bibnamefont
  {Kohama}}, \bibinfo {author} {\bibfnamefont {K.}~\bibnamefont {Iida}},\ and\
  \bibinfo {author} {\bibfnamefont {K.}~\bibnamefont {Oyamatsu}},\ }\href
  {https://doi.org/10.1103/PhysRevC.78.061601} {\bibfield  {journal} {\bibinfo
  {journal} {Phys. Rev. C}\ }\textbf {\bibinfo {volume} {78}},\ \bibinfo
  {pages} {061601} (\bibinfo {year} {2008})},\ \Eprint
  {https://arxiv.org/abs/0803.0187} {arXiv:0803.0187 [nucl-th]} \BibitemShut
  {NoStop}%
\bibitem [{\citenamefont {Rashdan}\ and\ \citenamefont
  {Sewailem}(2019)}]{Rashdan2019}%
  \BibitemOpen
  \bibfield  {author} {\bibinfo {author} {\bibfnamefont {M.}~\bibnamefont
  {Rashdan}}\ and\ \bibinfo {author} {\bibfnamefont {S.~M.}\ \bibnamefont
  {Sewailem}},\ }\href {https://doi.org/10.1142/S0218301319500149} {\bibfield
  {journal} {\bibinfo  {journal} {Int. J. Mod. Phys. E}\ }\textbf {\bibinfo
  {volume} {28}},\ \bibinfo {pages} {1950014} (\bibinfo {year}
  {2019})}\BibitemShut {NoStop}%
\bibitem [{\citenamefont {Liu}\ \emph {et~al.}(2003)\citenamefont {Liu},
  \citenamefont {Zhang},\ and\ \citenamefont {Zhang}}]{Liu2003PRC}%
  \BibitemOpen
  \bibfield  {author} {\bibinfo {author} {\bibfnamefont {Z.~H.}\ \bibnamefont
  {Liu}}, \bibinfo {author} {\bibfnamefont {X.~Z.}\ \bibnamefont {Zhang}},\
  and\ \bibinfo {author} {\bibfnamefont {H.~Q.}\ \bibnamefont {Zhang}},\ }\href
  {https://doi.org/10.1103/PhysRevC.68.024305} {\bibfield  {journal} {\bibinfo
  {journal} {Phys. Rev. C}\ }\textbf {\bibinfo {volume} {68}},\ \bibinfo
  {pages} {024305} (\bibinfo {year} {2003})}\BibitemShut {NoStop}%
\bibitem [{\citenamefont {Esbensen}(1996)}]{Esbensen1996PRC}%
  \BibitemOpen
  \bibfield  {author} {\bibinfo {author} {\bibfnamefont {H.}~\bibnamefont
  {Esbensen}},\ }\href {https://doi.org/10.1103/PhysRevC.53.2007} {\bibfield
  {journal} {\bibinfo  {journal} {Phys. Rev. C}\ }\textbf {\bibinfo {volume}
  {53}},\ \bibinfo {pages} {2007} (\bibinfo {year} {1996})}\BibitemShut
  {NoStop}%
\bibitem [{\citenamefont {Hebborn}\ and\ \citenamefont
  {Capel}(2019)}]{Hebborn2019PRC}%
  \BibitemOpen
  \bibfield  {author} {\bibinfo {author} {\bibfnamefont {C.}~\bibnamefont
  {Hebborn}}\ and\ \bibinfo {author} {\bibfnamefont {P.}~\bibnamefont
  {Capel}},\ }\href {https://doi.org/10.1103/PhysRevC.100.054607} {\bibfield
  {journal} {\bibinfo  {journal} {Phys. Rev. C}\ }\textbf {\bibinfo {volume}
  {100}},\ \bibinfo {pages} {054607} (\bibinfo {year} {2019})}\BibitemShut
  {NoStop}%
\bibitem [{\citenamefont {Kelley}\ \emph {et~al.}(1995)\citenamefont {Kelley},
  \citenamefont {Austin}, \citenamefont {Kryger}, \citenamefont {Morrissey},
  \citenamefont {Orr}, \citenamefont {Sherrill}, \citenamefont {Thoennessen},
  \citenamefont {Winfield}, \citenamefont {Winger},\ and\ \citenamefont
  {Young}}]{Kelley1995PRL}%
  \BibitemOpen
  \bibfield  {author} {\bibinfo {author} {\bibfnamefont {J.~H.}\ \bibnamefont
  {Kelley}}, \bibinfo {author} {\bibfnamefont {S.~M.}\ \bibnamefont {Austin}},
  \bibinfo {author} {\bibfnamefont {R.~A.}\ \bibnamefont {Kryger}}, \bibinfo
  {author} {\bibfnamefont {D.~J.}\ \bibnamefont {Morrissey}}, \bibinfo {author}
  {\bibfnamefont {N.~A.}\ \bibnamefont {Orr}}, \bibinfo {author} {\bibfnamefont
  {B.~M.}\ \bibnamefont {Sherrill}}, \bibinfo {author} {\bibfnamefont
  {M.}~\bibnamefont {Thoennessen}}, \bibinfo {author} {\bibfnamefont {J.~S.}\
  \bibnamefont {Winfield}}, \bibinfo {author} {\bibfnamefont {J.~A.}\
  \bibnamefont {Winger}},\ and\ \bibinfo {author} {\bibfnamefont {B.~M.}\
  \bibnamefont {Young}},\ }\href {https://doi.org/10.1103/PhysRevLett.74.30}
  {\bibfield  {journal} {\bibinfo  {journal} {Phys. Rev. Lett.}\ }\textbf
  {\bibinfo {volume} {74}},\ \bibinfo {pages} {30} (\bibinfo {year}
  {1995})}\BibitemShut {NoStop}%
\end{thebibliography}%

\end{document}